\documentclass{aa}

\usepackage{graphicx}
\usepackage{txfonts}

\usepackage{natbib}
\usepackage[colorlinks=true,linkcolor=blue,citecolor=blue]{hyperref}
\bibpunct{(}{)}{;}{a}{}{,}

\usepackage{multirow}

\usepackage{placeins}

\defcitealias{2024A&A...687A..45P}{\mbox{Paper~I}}
	
\title{Observationally derived change in the star formation rate as mergers progress\thanks{Table \ref{table:results:times} is only available in electronic form at the CDS via anonymous ftp to cdsarc.u-strasbg.fr (130.79.128.5) or via http://cdsweb.u-strasbg.fr/cgi-bin/qcat?J/A+A/}}

\author{W.~J.~Pearson\inst{\ref{inst:NCBJ}}
		\and L.~Wang\inst{\ref{inst:SRON}, \ref{inst:RUG}}
		\and V.~Rodriguez-Gomez\inst{\ref{inst:IdRA}}
		\and B.~Margalef-Bentabol\inst{\ref{inst:SRON}}
		\and L.~E.~Suelves\inst{\ref{inst:NCBJ}, \ref{inst:Tartu}}
}

\institute{National Centre for Nuclear Research, Pasteura 7, 02-093 Warszawa, Poland\label{inst:NCBJ}\\\email{william.pearson@ncbj.gov.pl}
	\and SRON Netherlands Institute for Space Research, Landleven 12, 9747 AD Groningen, The Netherlands\label{inst:SRON}
	\and Kapteyn Astronomical Institute, University of Groningen, Postbus 800, 9700 AV Groningen, The Netherlands\label{inst:RUG}
	\and Instituto de Radioastronom\'{i}a y Astrof\'{i}sica, Universidad Nacional Aut\'{o}noma de M\'{e}xico, Apdo. Postal 72-3, 58089 Morelia, Mexico\label{inst:IdRA}
	\and Tartu Observatory, University of Tartu, Observatooriumi 1, Tõravere 61602, Estonia\label{inst:Tartu}
}

\date{Received 29 November 2024; accepted 23 October 2025}

\abstract{Galaxy mergers can change the rate at which stars are formed. We can trace when these changes occur in simulations of galaxy mergers. However, for observed galaxies we do not know how the star formation rate (SFR) evolves along the merger sequence as it is difficult to probe the time before or after coalescence.}
{We aim to derive how SFR changes in observed mergers throughout the merger sequence, from a statistical perspective.}
{Merger times were estimated for observed galaxy mergers in the Kilo Degree Survey (KiDS) using a convolutional neural network (CNN). The CNN was trained on mock KiDS images created using IllustrisTNG data. The SFRs were derived from spectral energy density fitting to KiDS and VIKINGs data. To determine the change in SFR for the merging galaxies, each merging galaxy was matched and compared to ten comparable non-merging galaxies; matching each galaxy in redshift, stellar mass, and local density.}
{Mergers see an increase in the SFR for galaxies from 300~Myr before the merger until coalescence, continuing until at least 200~Myr after the merger event. After this, there is a possibility that SFR activity in the mergers begins to decrease, but we need more data to better constrain our merger times and SFRs to confirm this. We find that more galaxies with higher stellar mass (M$_{\star}$) have greater SFR enhancement as they merge compared to lower-M$_{\star}$ galaxies. There is no clear trend of changing SFR enhancement as local density changes, but the least dense environments have the least SFR enhancement. The increasing SFR enhancement is likely due to the closer proximity of galaxies and the presence of more close passes as the time before the merger approaches 0~Myr, with the SFR slowing 200~Myr after the merger event.}
{}


\begin{document}
\maketitle

\section{Introduction}\label{sect:intro}
In our current lambda cold dark matter paradigm, the dark matter halos in our Universe grow hierarchically through merging with one another. As the dark matter halos merge, the baryonic matter hosted within them also merges: the galaxies within the dark matter halos merge \citep[e.g.][]{2014ARA&A..52..291C, 2015ARA&A..53...51S}. These events, especially ones involving galaxies of a similar stellar mass (M$_{\star}$), can be violent and cause large disruption to the merging galaxies. Tidal tails can form between the merging galaxies, the interacting systems can become more irregular and asymmetric, ring structures can be created around the galaxies, and galaxies can be transformed from spirals to ellipticals \citep[e.g.][]{2013ApJ...778...61T, 2024MNRAS.529..810R}.

These morphological disturbances can be used to identify galaxy mergers and are exploited for visual identification, by both citizen scientists and professional astronomers \citep[e.g.][]{2008MNRAS.389.1179L, 2010MNRAS.401.1552D, 2010MNRAS.401.1043D, 2019AJ....158..103H, 2022A&A...661A..52P}. The parametric and non-parametric morphological parameters also change for merging galaxies, allowing parameters such as concentration, asymmetry, and smoothness \cite[CAS;][]{2000ApJ...529..886C, 2003AJ....126.1183C} as well as Gini and M$_{20}$ \citep{2004AJ....128..163L, 2008ApJ...672..177L, 2015MNRAS.451.4290S, 2019MNRAS.483.4140R} to be used to identify galaxy mergers. Morphological disturbances are not always needed: the close pairs method identifies galaxies that are close on the sky and moving slow enough, with respect to each other, to become gravitationally bound \citep[e.g.][]{2000ApJ...530..660B, 2005AJ....130.1516D, 2014MNRAS.444.3986R, 2018MNRAS.475.5133R, 2019ApJ...876..110D}.

More recently, advances in computational power have allowed galaxy mergers to be identified using machine learning techniques. Convolutional neural networks (CNNs) have been used to replicate visual identification \citep[e.g.][]{2018MNRAS.479..415A, 2019MNRAS.490.5390B, 2019A&A...631A..51P, 2022A&A...661A..52P, 2019MNRAS.483.2968W, 2020A&C....3200390C, 2021MNRAS.506..677C, 2020A&A...644A..87W, 2021MNRAS.504..372B, 2022MNRAS.514.3294B}. Much like traditional visual classification, this has relied on identifying morphological disturbances and faint features of the merger to produce reliable merger identifications \citep[e.g.][]{2019A&A...626A..49P, 2022A&A...661A..52P}. Machine learning methods can also be trained with morphological parameters of galaxies \citep{2019MNRAS.486.3702S, 2022A&A...661A..52P, 2023MNRAS.519.4920G, 2024A&A...687A..24M} and photometry \citep{2023A&A...669A.141S}.

The different methods of identifying galaxy mergers identify different stages of mergers. The close pairs method typically identifies pre-mergers while morphological parameters are good at identifying post-mergers; galaxies that are identified with both methods are then typically found to be ongoing mergers \citep{2023MNRAS.523.4381D, 2024MNRAS.528.5558W}. Machine learning methods trained with galaxy mergers identified with these methods are likely to similarly select pre- or post-mergers. Studies have also used machine learning techniques to specifically identify post-merging galaxies \citep{2021MNRAS.504..372B, 2022MNRAS.514.3294B, 2024MNRAS.528.5558W}. Identifying non-mergers, pre-mergers, and post-mergers with machine learning has also been studied, although with less effectiveness than non-merger and merger identification \citep{2020ApJ...895..115F, 2024A&A...687A..24M}.

Further refinement to determine the precise time before or after a merger has also been conducted \citep{2021arXiv210205182K, 2024A&A...687A..45P}. Here, merging galaxies from cosmological simulations with known times before or after a merger event are used to train deep neural networks. Images of these galaxies have been combined with their physical properties \citep{2021arXiv210205182K} or used on their own \citep[][hereafter Paper I]{2024A&A...687A..45P}. These approaches do not attempt to identify galaxy mergers, only determine the merger time for pre-identified merging galaxies. The merger time prediction only for post-merger galaxies has also been conducted with coarse time resolution, again using simulations to train deep learning networks (Ferreira et al. in prep).

The interactions of the galaxies, and their morphological changes, drive material around within the galaxies themselves. If this movement of dust and gas is driven into the centre of a galaxy, it can trigger a merger-induced active galactic nucleus \citep[AGN;][]{1985AJ.....90..708K, 2011ApJ...743....2S, 2012A&A...538A..15H, 2017MNRAS.464.3882W, 2019MNRAS.487.2491E, 2020A&A...637A..94G, 2021ApJ...909..124S, 2023MNRAS.519.6149B, 2024A&A...690A.326L}. It is also seen to decrease the metallicity of the gas in the centre of galaxies \citep{2008AJ....135.1877E, 2010ApJ...710L.156R, 2012MNRAS.426..549S, 2018MNRAS.479.3381B, 2017MNRAS.467.3898C, 2022MNRAS.509.2720S}.

Tidal forces created by galaxy mergers can also cause the gas in the interacting galaxies to be compressed and shocked. This can result in an increased star formation rate (SFR) in merging galaxies, in some cases to such a degree that they be considered to be starbursts \citep[e.g.][]{2008AJ....135.1877E, 2011A&A...535A..60H, 2012MNRAS.426..549S, 2013MNRAS.435.3627E, 2013MNRAS.433L..59P, 2022MNRAS.516.4922R, 2024MNRAS.52711372A}. It has been shown that the higher a galaxy's SFR is, at fixed M$_{\star}$, the more likely it is to be undergoing a merger \citep{2019A&A...631A..51P}. However, the typical increase in the SFR caused by mergers is low, approximately a factor of two, and some works argue that there is no increase in the SFR during merger events \citep{2015MNRAS.454.1742K, 2022ApJ...940....4S}. This low, or lack of, SFR enhancement may be a result of the environment in which the merging galaxies lie, with galaxy mergers in higher-density environments found to have lower SFR enhancement \citep{2024RAA....24e5005H}.

This difference in SFR enhancement found in different studies may be a result of different research using different selections of galaxy mergers. Studies using pair-selected galaxy mergers, which are pre-mergers, find correlations between the SFR and the separation between the galaxies: smaller separations have a higher SFR \citep{2013MNRAS.433L..59P, 2022ApJ...940....4S, 2023MNRAS.522.5107B}. Although, again, there is evidence of a reduced SFR in close pairs where one galaxy is elliptical with this suppression stronger for smaller galaxy separations \citep{2024ApJ...965...60F}. The SFR enhancement in post-mergers is consistent with SFR enhancement of the closest pre-mergers \citep{2022MNRAS.514.3294B}. \citet{2022MNRAS.514.3294B} also found that if their neural-network-selected post-mergers have a larger SFR enhancement, the galaxy is more likely to be a recent post-merger, possibly indicating a reduction in SFR enhancement with time for post-merger galaxies. Post-merger galaxies are found to have younger stellar populations compared to non-mergers \citep{2024MNRAS.527.2037R}. There is also evidence that post-mergers undergo a rapid SFR decline after their recent burst of star formation \citep{2022MNRAS.517L..92E}.

This apparent link between SFR enhancement and separation distance for pre-mergers as well as SFR enhancement and confidence in identifying post-mergers suggests there is a temporal link between SFR enhancement and the time before or after a merger event. Indeed, this is seen in simulations. Post-merger galaxies in the IllustrisTNG simulations \citep{2017MNRAS.465.3291W, 2018MNRAS.480.5113M, 2018MNRAS.477.1206N, 2018MNRAS.475..624N, 2018MNRAS.475..648P, 2018MNRAS.473.4077P, 2018MNRAS.475..676S} have a decreasing SFR as the time since merger increases \citep{2020MNRAS.493.3716H}. Merging galaxies in the FIRE-2 zoom-in simulations \citep{2018MNRAS.480..800H} have increased SFR at first and second peri-centric passages, increased SFR enhancement as the time to coalescence decreases, followed by a decrease in the SFR as the time after the merger event increases \citep{2019MNRAS.485.1320M}. Such a study is not trivial with observations when considering both pre-mergers and post-mergers. Due to the merger timescale of over a billion years, we cannot watch real galaxies merge and trace their SFR enhancement through the pre-merger and post-merger phases. The time before or after a merger event is difficult to determine and has only been done twice \citep{2021arXiv210205182K, 2024A&A...687A..45P}, and primarily for simulated galaxies.

In this work we aim to produce a statistical understanding of how the the SFR changes as a merger progresses. We trained a neural network to determine the time before or after a merger event for merging galaxies using images and merger times from IllustrisTNG simulations. This network was trained such that is is applicable to data from the Kilo Degree Survey \citep[KiDS;][]{2013Msngr.154...44D, 2013ExA....35...25D}. From this, generated a statistical SFR enhancement as a function of time before and after a merger event. This was done by comparing merging galaxies to non-merging galaxies where the primary difference was if a galaxy was merging or not: we assumed that the merging galaxies were randomly sampled from similar non-merging galaxies.

The paper is structured as follows. Section \ref{sect:data} describes the IllustrisTNG and KiDS data and Sect. \ref{sect:nn} describes our neural network. Our results are presented and discussed in Sects. \ref{sect:results} and \ref{sect:discussion}, respectively. We conclude in Sect. \ref{sect:conclusion}. Where necessary, we follow the Planck2015 cosmology \citep{2016A&A...594A..13P}.

\section{Data}\label{sect:data}
\subsection{IllustrisTNG}
The time before or after a merger event is difficult to determine for real galaxies. However, these times are better known for galaxies in simulations, although the precision is subject to the temporal resolution of the simulation. In this work, we use galaxy mergers from IllustrisTNG 100-1, with known times before or after a merger event (merger time), to train our neural network.

\subsubsection{Simulation details}
IllustrisTNG's TNG100-1 is the high-resolution run of the TNG100 series of simulations and simulates both dark and baryonic matter. TNG100-1 simulates 1820$^{3}$ dark matter particles within a box of 75\,000~ckpc/h per side, and uses Planck2015 cosmology. The dark matter particles in this simulation have a mass of $\approx 5\times10^{6}$~M$_{\odot}$/h and the gas cells have an average mass of $\approx 9\times10^{5}$~M$_{\odot}$/h.

The simulation begins at $z=127$ and end at $z=0$, with the time between snapshots varying between 30~Myr and 234~Myr; larger times between snapshots are typically found at lower redshifts. The galaxies found in IllustrisTNG are found to be morphologically similar to observed galaxies; relations between M$_{\star}$ and morphologies, shapes, and sizes are found be be within $1\sigma$ of the observed trends \citep{2019MNRAS.483.4140R}. On the other hand, at fixed M$_{\star}$, spiral galaxies are not found to be larger than elliptical galaxies, unlike observed galaxies, and the morphology-colour relation is not strong \citep{2019MNRAS.483.4140R}.

\subsubsection{Mergers}
To select the major merging galaxies from this simulation, we followed \citetalias{2024A&A...687A..45P}. A merging galaxy is initially defined as a galaxy that has merged in the last 500~Myr (post-merger) or will merge in the next 1~Gyr (pre-merger). Throughout this paper, the `merger time' can be considered to be the time after the merger event. Thus, a merger time of 0~Myr is the merger event, positive merger times are post-mergers, and negative merger times are pre-mergers. These times are initially derived from the time between the snapshot that a galaxy is `observed' in the simulation and the `snapshot of merger'. The snapshot of merger is the snapshot where multiple galaxies in earlier snapshots are tracked as a single galaxy for the first time using the Sublink\_gal merger trees \citep{2005Natur.435..629S, 2009MNRAS.398.1150B, 2015MNRAS.449...49R}.
We define a major merger in the simulation to be a merger between at least two galaxies where the ratio between the stellar masses of the two most massive galaxies is $<1:4$ (i.e. the larger galaxy has M$_{\star}$ no more than four times larger than the M$_{\star}$ of the smaller galaxy) in the snapshot where the second most massive galaxy reached its maximum M$_{\star}$ \citep{2015MNRAS.449...49R}. We also required that the baryonic component (stars + star-forming gas) of the merging galaxy's subhalo be at least $1\times10^{9}$~M$_{\odot}$/h.

\subsubsection{Merger sample and times}
As with \citetalias{2024A&A...687A..45P}, we selected merging galaxies from snapshots 87 to 93, inclusive, which correspond to redshifts between 0.07 and 0.15. The upper limit was selected to match the upper redshift limit of our KiDS merging galaxy catalogue \citep{2019A&A...631A..51P}. The lower limit allows the full 1~Gyr pre-merger requirement to elapse before $z = 0$ in the IllustrisTNG simulations, allowing a more complete pre-merger sample. If a galaxy is found to be a merger at any merger stage in more than one snapshot, the same galaxy will be selected in each snapshot that it is found to be a merger. This sample selection resulted in 6139 major mergers from TNG100-1.

To increase the precision of the merger times, simple gravity simulations were applied to each major merger selected, again following \citetalias{2024A&A...687A..45P}. Here, each galaxy with a mass ratio $<1:10$ of the most massive galaxy in the merger was treated as a point mass in 3D space and allowed to move only under the gravity of the other galaxies in the merger. Galaxies with a mass ratio $<1:10$ were included to allow the smaller galaxies to influence the merger time. A softening length of 0.5~ckpc/h was used to match TNG100-1. The simulation was run between the snapshot of merger and the previous snapshot with a time-step of 0.001~Myr. The merger was considered complete when the major galaxies were first at their closest approach. If a merger contains more than two major galaxies, the merger was considered complete the first time the first two major galaxies are at their closest for pre-mergers and the first time the last two major galaxies are at their closest for post-mergers. The resulting merger times were rounded to the nearest million years. We note that as these simple simulations are only considering gravity, they do not accurately reproduce the final positions of the galaxies in the IllustrisTNG simulations. If the simple gravity simulations did not produce a closest approach, the merger was removed from further study. We also removed galaxies that are classified as pre-mergers and as post-mergers to avoid ambiguity. This further refinement of our sample selection left 3321 major mergers from TNG100-1.

After initial testing, we found that using the full 1~Gyr pre-merger time range produces poor results. This is consistent with literature in which mergers more than 500~Myr before a merger event are shown to be difficult to identify and classify \citep{2025A&A...697A.207D}. Instead, we reduced the pre-merger time scale to 500~Myr. The left a sample of 2213 major mergers from TNG100-1 that we used for further analysis.

\subsubsection{Merger images}
Four channel images were created of the 2213 major mergers from the IllustrisTNG simulation. To do this, each stellar particle had a spectral energy distribution derived from the stellar properties of the particle and its corresponding \citet{2003MNRAS.344.1000B} stellar population model. The resulting spectra were passed through the KiDS $u$, $g$, $r$, and $i$ filters to create a smoothed 2D image in each band \citep{2019MNRAS.483.4140R}. We matched the pixel resolution of the IllustrisTNG images to that of KiDS, 0.2~arcsec/pixel, and create four channel images of $128\times128$ pixels centred on the merging galaxy. The images of the galaxies were then made such that the galaxies appears to lie at the redshift of the snapshot the galaxy was selected from. This means that the physical to angular size relation and dimming are correct for the redshift of the snapshot of each galaxy. As with \citetalias{2024A&A...687A..45P}, we do not require the companion galaxy of the pre-mergers to be within the image. The application of radiative transfer to simulated galaxy images has been shown to not cause large changes in galaxy morphology \citep{2019MNRAS.483.4140R} or notably change the performance of neural networks in classifying galaxy mergers \citep{2019MNRAS.490.5390B}. Therefore, we elect not to calculate radiative transfer on our sample of IllustrisTNG galaxies, saving computational time. Three images of each merging galaxy were created, viewed from along the x, y, and z axes of the IllustrisTNG simulations. This increases the number of merging galaxy images by a factor of three, providing 6639 images for further use.

\begin{figure}
	\resizebox{\hsize}{!}{\includegraphics{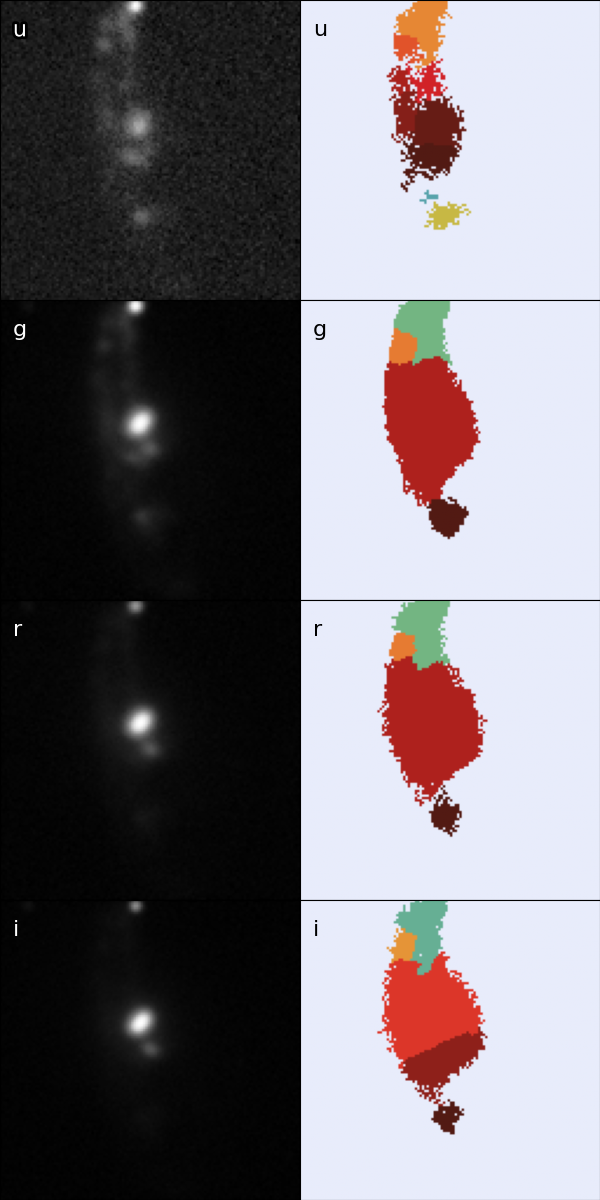}}
	\caption{Randomly selected example IllustrisTNG image after processing to appear like a KiDS image. The target galaxy is in the centre of the image. The left column shows the $u$, $g$, $r$, and $i$ band images, as labelled, with arcsinh scaling applied twice. The right column shows the associated segmentation maps for each band, with each colour representing a different segment. Each panel is $25.6 \times 25.6$~arcsec ($128 \times 128$ pixels), corresponding to a physical size of $53 \times 53$~kpc at the redshift of the IllustrisTNG galaxy ($z=0.11$).}
	\label{fig:data:tngexample}
\end{figure}

To train a neural network to be applied to real KiDS images, the IllustrisTNG images of simulated galaxies need to be processed to resemble real KiDS observations as closely as possible \citep[e.g.][]{2019MNRAS.490.5390B, 2019A&A...626A..49P, 2020arXiv201103591C, 2023MNRAS.521.3861D}. Within each KiDS survey tile that has a source catalogue, we determined the average separation between the bright KiDS sources \citep[$r$-mag $< 20$;][]{2021A&A...653A..82B} at $z < 0.3$. Random positions that are at least the average separation away from the bright sources at $z < 0.3$ were found. Using these random positions, 500 cut-outs that were $128\times128$ pixels in size and did not contain masked pixels in any KiDS band were made in each of the 1001 KiDS tiles. This resulted in 500\,500, $ugri$ cut-outs being created. This allows background galaxies in the cut-outs but avoids bright sources at similar redshifts to our IllustrisTNG galaxies. For each IllustrisTNG image, we randomly select a cut-out, convolve each band of the IllustrisTNG images with their respective point spread function (PSF) of the tile the cut-out was made in, and place the IllustrisTNG galaxy into the KiDS cut-out to create the mock $ugri$ KiDS images. As we require the PSF, we only use cut-outs that have a well defined PSF in all four $u$, $g$, $r$, and $i$ KiDS bands. We then created segmentation maps for each band of the mock-KiDS images using \texttt{photutils} \citep{larry_bradley_2024_10671725}. If sources could not be found in the segmentation map with \texttt{photutils}, the pixel values of the IllustrisTNG images, without KiDS noise or PSF convolution, were increased by a factor of two, and the process to create a KiDS-like image repeated. If this IllustrisTNG image scaling was done four times and creation of the segmentation map still failed, the merger image was rejected. Scaling the images in this way effectively increases the brightness of the galaxies, increasing the signal-to-noise, but does not bias the training. This left all 6639 KiDS-like IllustrisTNG images. An example of a merging IllustrisTNG galaxy, and its segmentation maps, can be found in Fig. \ref{fig:data:tngexample}.

\subsection{KiDS}
KiDS is a wide-field, multi-band survey performed with the OmegaCAM of the VLT Survey Telescope \citep{2011Msngr.146....2C, 2011Msngr.146....8K, 2012sngi.confE...1C}. It covers over 1000 square degrees in the $u$, $g$, $r$, and $i$ bands and covers the same survey area as the VISTA Kilo-degree Infrared Galaxy survey \citep[VIKING;][]{2013Msngr.154...32E}. In this work, we use the latest KiDS data release 4 \citep[DR4;][]{2019A&A...625A...2K}. These data have average 5$\sigma$ limiting magnitudes of 24.13, 25.12, 25.02, and 23.68~Mag and average PSF full-width at half-maximum of 0.58, 0.54, 0.46, and 0.49~arcsec for the $u$, $g$, $r$, and $i$ bands, respectively.

\subsubsection{Merger and control selection}\label{sect:data:mergers}
We identified galaxies in KiDS by using the star-galaxy flags provided in DR4. We required that galaxies have a \texttt{CLASS\_STAR} value less than or equal to 0.05, \texttt{SG2DPHOT} value of 0, and a \texttt{SG\_FLAG} of 1. We also required that all galaxies have 5$\sigma$ detections in the $u$, $g$, $r$, and $i$ bands. We selected galaxy mergers from these KiDS galaxies by selecting all KiDS galaxies where both the peak of the photometric redshift posterior (z$_{B}$) and the maximum likelihood redshift (z$_{ML}$) are less than 0.15. We also require that a $256\times256$~pixel, $r$-band cut-out can be made without filling empty space (i.e. the galaxy's position must be more than 128~pixels from the edge of the $r$-band image). The resulting 409\,934 galaxies were then passed through the CNN of \citet{2019A&A...631A..51P}. Briefly, this network was trained to identify galaxy mergers in KiDS $r$-band images using galaxies identified as merging by the GAMA-KiDS Galaxy Zoo project \citep{2024PASA...41..115H} as well as being loosely identified as merging via a loose $r$-band asymmetry-smoothness cut where mergers lie above $A = 0.35S + 0.02$ \citep{2003AJ....126.1183C}. The \citet{2019A&A...631A..51P} network finds 97\,891 galaxy merger candidates.

The galaxy merger sample was likely to be highly contaminated \citep[e.g.][]{2021MNRAS.504..372B, 2022A&A...661A..52P}. As the size of the merger sample is too large to be able to be visually checked, we elected to select merging galaxies where the \citet{2019A&A...631A..51P} catalogue's merger probability is greater than 0.99; that is, the galaxies that the neural network is the most confident are merging. We also limited the sample to redshifts between 0.07 and 0.15 and log(M$_{\star}$) $>$ 9, to ensure our sample is above the lower mass limit of the IllustrisTNG training data. This M$_{\star}$ limit is also approximately 0.8~dex above the expected KiDS-VIKING mass completeness limit of log(M$_{\star}$) $\approx$ 8.2 at z = 0.15 \citep{2019A&A...632A..34W}. This results in a sample of 10\,448 merging KiDS galaxies.

To be able to determine the change in the SFR due to galaxy mergers, we required a control sample of non-merging galaxies. We selected non-merging galaxies to be those where the \citet{2019A&A...631A..51P} catalogue's merger probability is less than 0.05, providing 240\,129 non-merging KiDS galaxies. Each merging galaxy had ten non-merging control galaxies selected, with replacement, which match closely in redshift, local density, and M$_{\star}$ \citep{2013MNRAS.430.3128E}. We assume the merging galaxies are randomly sampled from the entire galaxy population where these physical properties are similar. These non-merging galaxies were required to have redshifts within 0.02, environmental density (Sect. \ref{sect:data:density}) within 0.02, and log(M$_{\star}$) within 0.02 of the merging galaxies. Of our merger sample, 10\,029 merging galaxies could be matched to 10 unique non-merging galaxies, and were used for further study. If more than 10 non-mergers matched to a merging galaxy, 10 matching non-mergers were randomly selected.

\begin{figure}
	\resizebox{\hsize}{!}{\includegraphics{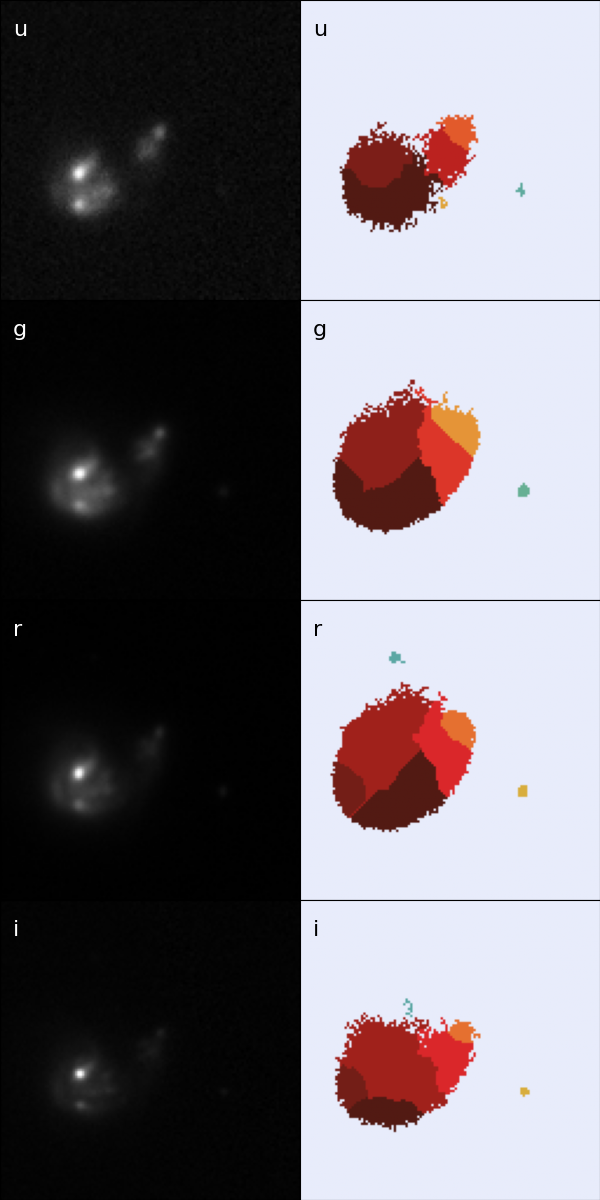}}
	\caption{Randomly selected example KiDS image of a merging galaxy. The target galaxy is in the centre of the image. The left column shows the $u$, $g$, $r$, and $i$ band images, as labelled, with arcsinh scaling applied twice. The right column shows the associated segmentation maps for each band, with each colour representing a different segment. Each panel is $25.6 \times 25.6$~arcsec ($128 \times 128$ pixels), corresponding to a physical size of $44 \times 44$~kpc at the redshift of the galaxy ($z=0.09$).}
	\label{fig:data:kidsexample}
\end{figure}

The merger sample was further reduced to a subsample of galaxies for which 128$\times$128 pixel cut-outs can be made in all four KiDS bands without filling empty space. These cut-outs could not contain any masked $r$-band pixels. We created segmentation maps of these images, again using \texttt{photutils}, and rejected galaxies that could not be detected in the segmentation map or whose segmentation map caused errors in \texttt{photutils} when we attempted to deblend it into its components. This leaves 5897 merging galaxies whose sources can be detected and who have at least 10 non-merging control matches. An example of a merging KiDS galaxy, and its segmentation maps, can be found in Fig. \ref{fig:data:kidsexample}.

\subsubsection{Environmental density}\label{sect:data:density}
To determine the environmental density, we followed \citet{2021MNRAS.507.3070S} and used the density relative to the distance of the tenth nearest neighbour:
\begin{equation}\label{eqn:density}
	\Sigma_{10} = \frac{10}{\theta_{10}^{2}},
\end{equation}
where $\theta_{10}$ is the angular projected distance to the tenth nearest neighbour, in megaparcecs. The tenth nearest neighbour was determined within a redshift bin centred on the maximum likelihood redshift ($z_{ML}$) of the object being studied and with half width ($\sigma_{bin}$) determined by the uncertainty in the photo-z estimate. That is:
\begin{equation}
	\sigma_{bin} = \frac{(z_{B, max} - z_{B, min})*(1+z_{ML})}{2},
\end{equation}
where $z_{B, max}$ and $z_{B, min}$ are the upper and lower bounds of the 68\% confidence interval of $z_{B}$.

Galaxies close to the edges of a survey will have an underestimated local density due to the missed galaxies outside of the survey area. To correct for this, we again followed \citet{2021MNRAS.507.3070S}. Edge galaxies were selected where $\theta_{10}$ is larger than the distance to the edge of the survey. A circular area with radius $\theta_{10}$ was calculated (A$_{10}$) and the fraction of this area that is outside the survey ($x$) determined. Thus, the area (1-$x$)A$_{10}$ contains ten galaxies. If uniform density is then assumed, we can calculate the number of galaxies ($n$) that should be in the survey coverage of A$_{10}$ if $\theta_{10}$ was the distance to the true tenth nearest neighbour as:
\begin{equation}
	n = 10(1-$x$).
\end{equation}
The true $\theta_{10}$ can then be estimated by calculating $n$, rounding to the nearest integer, and taking the $n$th nearest neighbour projected distance to be $\theta_{10}$. The robustness of this method is discussed in \citet[][Appendix B]{2021MNRAS.507.3070S}.

\subsubsection{Star formation rates}\label{sect:data:sfr}
The SFR (and M$_{\star}$) of the KiDS galaxies were derived using Code Investigating GALaxy Emission \citep[CIGALE;][]{2019A&A...622A.103B}. For this, we used an exponentially declining star formation history with an exponential burst and \citet{2003MNRAS.344.1000B} stellar population model with the \citet{2003PASP..115..763C} initial mass function. For the dust, we used a modified \citet{2000ApJ...539..718C} attenuation model and the \citet{2014ApJ...780..172D} emission model. We also include the \citet{2006MNRAS.366..767F} AGN model. The full setup can be found in Appendix \ref{app:cigale}.

We used CIGALE to fit these models to the four KiDS optical bands ($u$, $g$, $r$, and $i$) and the five VISTA Kilo-degree Infrared Galaxy Public Survey \citep[VIKING;][]{2013Msngr.154...32E} infrared bands ($z$, $y$, $J$, $H$, and $Ks$). While far-infrared data are useful to constrain the SFR, these data are only available in parts of the Northern KiDS field from the \textit{Herschel} atlas \citep{2010PASP..122..499E, 2019MNRAS.490..634S}. To not limit our sample, we elect to not require far-infrared data with the understanding that the SFRs of dusty galaxies could be particularly unreliable. The mean (median) error on our log(SFR/M$_{\star}$ yr$^{-1}$) estimates is 0.61 (0.61). Further discussion on our SFR (and M$_{\star}$) values can be found in Appendix \ref{app:sfr}.

\section{Neural network}\label{sect:nn}
In this work, we trained a neural network with IllustrisTNG data to predict the merger time that will be applied to KiDS images. Generally, this domain switching cannot be done with machine learning unless the tool has been trained to work in the target domain. Hence the work performed to make the IllustrisTNG images look like KiDS images. Our neural network uses the best performing architecture from \citetalias{2024A&A...687A..45P}, a CNN. In short, the network begins with six convolutional layers, each using a stride of 1 and `same' padding, with each convolutional layer followed by batch normalisation, dropout with a rate of 0.2, and $2\times2$ max pooling. These convolutional layers have 32. 64, 128, 256, and 1024 square filters with 6, 5, 3, 3, and 2 pixels on a side. The convolutional layers are followed by fully connected layers of 2048, 512, and 128 neurons. Each fully connected layer is followed by batch normalisation and dropout with a rate of 0.1. All layers in the feature extractor and classifier use Rectified Linear Units \citep[ReLU;][]{Nair2010} activation.

The input to the network is an eight channel, 128$\times$128 pixel image. Four channels are the $u$, $g$, $r$, and $i$ band images while the other four are segmentation maps, one for each of the bands. Tests were performed training the network with all 15 combinations of the four bands (image plus segmentation map), and using $u$, $g$, $r$, and $i$ proved to be the most reliable. Following \citetalias{2024A&A...687A..45P}, the image channels are arcsinh scaled twice and then normalised between 0 and 1. The segmentation maps are linearly scaled between 0 and 1. The output is a single neuron with sigmoid activation, where the output is a linear scaling of the merger time: 0 is 500~Myr and 1 is $-500$~Myr.

The IllustrisTNG merger sample was split into three groups: training with 5286 images (1762 galaxies), validation with 672 images (224 galaxies), and test with 681 images (227 galaxies). The training data was used to train the CNN while the validation was used to trace how well the CNN was functioning during training. The test set was used once, and only once, after training was completed to provide information on how well the CNN was performing on unseen data. We enforced that a merger tree cannot be split between the training, validation, and test sets so galaxies from one merger tree are only found in one of the three sets. This prevents the quality of the CNN from being overestimated due to the network learning to interpolate along a merger tree in a way that does not apply to unseen data.

The training was performed using the normalised merger time, calculated as
\begin{equation}\label{eqn:nn:normalise}
	\mathrm{normalised~merger~time} = -1\times\frac{(\mathrm{merger~time / Myr})+500}{1000},
\end{equation}
meaning normalised merger times of 0 and 1 correspond to merger times of 500~Myr and $-500$~Myr, respectively. The true normalised merger times were compared to the predicted merger times using mean squared error (MSE) loss. We used the Adam optimiser \citep{2014arXiv1412.6980K} with an initial learning rate of 5e-5 and ran training for 1000 epochs. The code to train the network and the trained model are available on GitHub\footnote{\url{https://github.com/wjpearson/KiDSMergerTime}}.

\section{Results}\label{sect:results}
\subsection{Neural network}\label{sect:results:nn}
\begin{figure}
	\resizebox{\hsize}{!}{\includegraphics{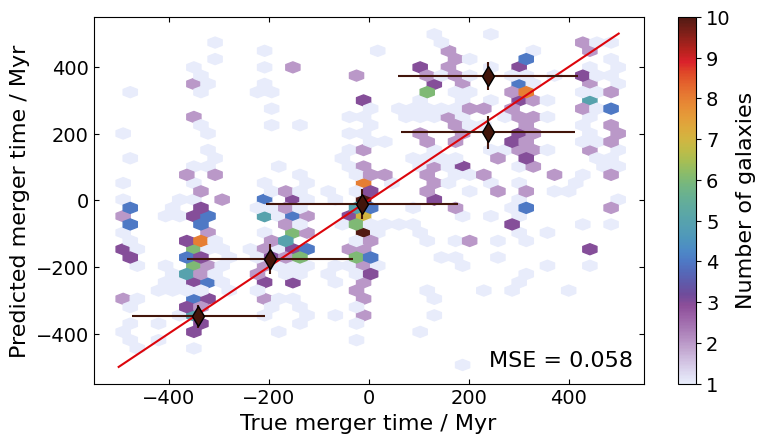}}
	\caption{Predicted merger times of the test data against the true merger times. The colour corresponds to the number density from low (purple) to high (red). The red line indicates a one-to-one relation and the MSE for the normalised times is shown in the bottom right. Dark brown diamonds show the median true merger times and median predicted merger times in predicted merger time bins with widths of 200~Myr; x errors are the median absolute deviation of the true times in the bin and y errors are the median absolute deviation of the predicted times.	
	}
	\label{fig:result:test}
\end{figure}

The final CNN, chosen as the epoch with the lowest validation loss, has a MSE of 0.067. When then applied to the previously unseen test data, the MSE reduced slightly to 0.058. The predicted merger times against the true merger times of the test data set can be found in Fig. \ref{fig:result:test}. The test result merger time had a mean difference between the true and predicted merger times (mean error) of 182~Myr and a median difference between the true and predicted merger times (median error) of 145~Myr.

We binned the predicted merger times into bins of width 200~Myr and present the median true merger times and median predicted merger times in Fig. \ref{fig:result:test} as dark brown diamonds. The true time for the mergers with predicted merger time $>$ 0~Myr tend towards 240~Myr, close to the mid of the post-merger time range (250~Myr). The mergers with a true merger time $<$ 0~Myr appear to be reasonably well reproduced. Therefore, the CNN is able to reasonably assign merger times to the pre-merger galaxies, but struggles with post-merger and assigns them merger times centred close to the median post-merger time. We also calculated the mean and median error of the predicted merger times in each bin. These errors are presented in Table \ref{table:results:errors}.
	
\begin{table}
	\caption{Mean and median predicted merger time errors within predicted merger time bins.}
	\label{table:results:errors}
	\centering
	\begin{tabular}{ccc}
		\hline
		\hline
		Predicted merger & Mean error & Median error \\
		time ($t$) range & (Myr) & (Myr) \\
		\hline
		-500~Myr $\leq t <$ 500~Myr & 182 & 145 \\
		-500~Myr $\leq t <$ -300~Myr & 140 & 60 \\
		-300~Myr $\leq t <$ -100~Myr & 165 & 145 \\
		-100~Myr $\leq t <$ 100~Myr & 190 & 160 \\
		100~Myr $\leq t <$ 300~Myr & 179 & 154 \\
		300~Myr $\leq t <$ 500~Myr & 214 & 154 \\
		\hline
	\end{tabular}
\end{table}

\subsection{KiDS merger time}\label{sect:results:kids}
\begin{figure}
	\resizebox{\hsize}{!}{\includegraphics{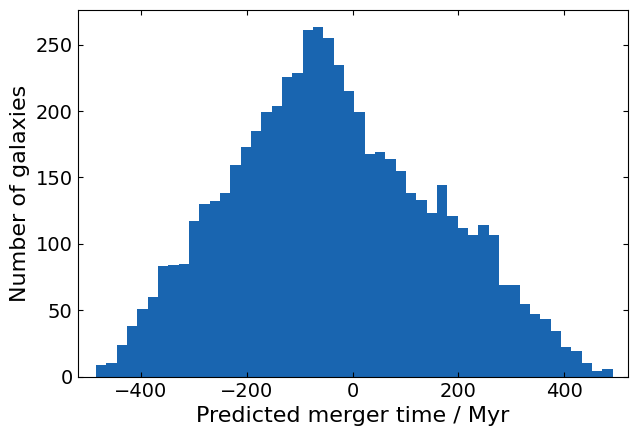}}
	\caption{Distribution of predicted merger times for KiDS merging galaxies. Most of the KiDS mergers are pre-mergers, with merger times less than 0~Myr.}
	\label{fig:results:times}
\end{figure}

\begin{figure*}
	\centering
	\resizebox{0.9\hsize}{!}{\includegraphics{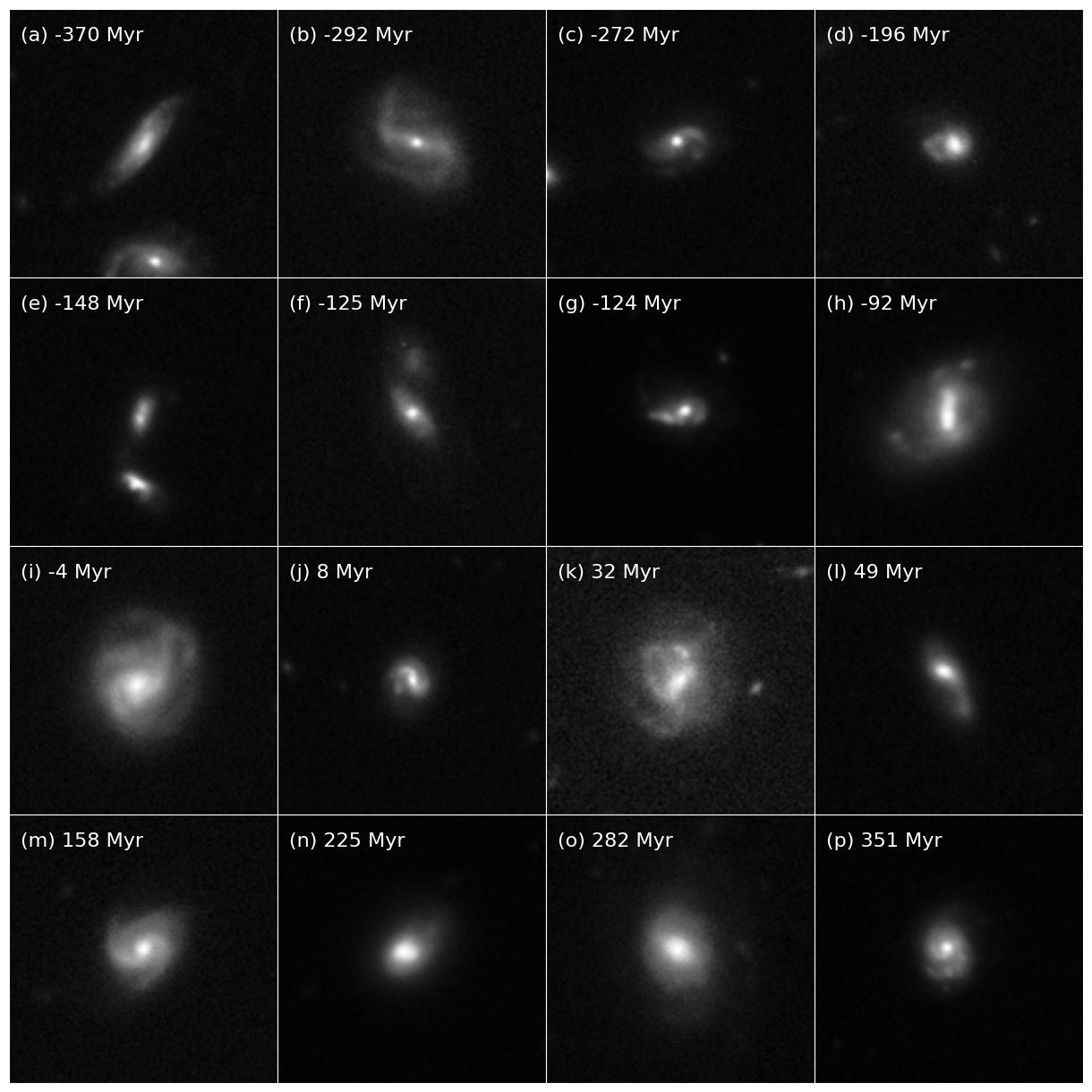}}
	\caption{Sixteen KiDS merging galaxies with their associated predicted merger times (time after a merger). The shown images are $r$-band images with arcsinh scaling applied twice. The galaxy in the centre of the image is the galaxy whose merger time is shown. Each panel is $25.6 \times 25.6$~arcsec ($128 \times 128$ pixels), corresponding to a physical size of $53 \times 53$~kpc at the median redshift of the KiDS merger sample ($z=0.11$).}
	\label{fig:results:examples}
\end{figure*}

We applied our neural network to our sample of 5897 merging galaxies described in Sect. \ref{sect:data:mergers}. We present these merger times in Table \ref{table:results:times} and their distribution in Fig. \ref{fig:results:times}. We note that the merger times can be calculated from the normalised merger times predicted by the neural network by inverting Eq. \ref{eqn:nn:normalise}; that is to say,
\begin{equation}\label{eqn:results:time}
	\mathrm{merger~time} / Myr = -1\times(\mathrm{normalised~merger~time} \times 1000)-500.
\end{equation}
The majority of our mergers have merger times less than zero, implying our KiDS mergers are primarily pre-mergers. Examples of KiDS mergers with their predicted merger times can be found in Fig. \ref{fig:results:examples}.

\begin{table}
	\caption{Predicted merger times for KiDS merging galaxies}
	\label{table:results:times}
	\centering
	\begin{tabular}{ccc}
		\hline
		\hline
		\multirow{ 2}{*}{KiDSDR4 ID} & Merger time & Merger time\tablefootmark{a} \\
		& Normalised & Myr \\
		\hline
		J000629.821-334331.72 & 0.217 & 283 \\
		J032443.393-271113.19 & 0.528 & -28 \\
		J114915.673-005836.62 & 0.642 & -142 \\
		J084935.741-014354.00 & 0.381 & 119 \\
		J152843.103+011926.18 & 0.422 & 78 \\
		J143757.686-003357.28 & 0.589 & -89 \\
		J222645.424-274612.62 & 0.635 & -135 \\
		J085728.553-005509.55 & 0.549 & -49 \\
		J103131.827-022242.17 & 0.72 & -220 \\
		J000920.447-285531.03 & 0.225 & 275 \\
		... & ... & ... \\
		\hline
	\end{tabular}
	\tablefoot{\tablefoottext{a}{Merger times calculated using Eq. \ref{eqn:results:time}.}\\
		The full table is available at the CDS.}
\end{table}

In Fig. \ref{fig:results:examples}, panels a to i show mergers with predicted merger times $<$ 0~Myr. Panel i is the closest to coalescence, only 4~Myr before, has large disturbances but does not show a secondary nuclei. This may suggest that this galaxy is actually recent post merger. All the remaining panels, except panel b, show evidence of a double nuclei or a close companion indicating that these are all pre-mergers. Where a secondary galaxy is present, in panels a, c, e, f, and g, there is no clear trend of decreasing separation with decreasing merger time. This would indicate that we see mergers with different approach speeds or galaxies potentially before and after a close passage. However, the lack of large disturbances for galaxies with smaller merger times but larger separations suggests that we are not primarily seeing galaxies after a close passage in this figure. Panel b has no clear secondary galaxy but show clear distortions in the spiral arms. It is likely that this merger has a predicted merger time that is under-estimated, as it appears similar to what we would expect a merger with positive merger time to look like. It is also possible that the secondary galaxy is outside the cut-out or hidden directly behind the primary galaxy.

The remaining panels in Fig. \ref{fig:results:examples} show mergers with predicted merger times $>$ 0~Myr. These galaxies all show signs of disturbance but lack a companion or obvious, separated double-nuclei.  All seven galaxies, however, show signs of disturbance which appears to be reduced as the time since merger increases. The exception is panel j, that while has disturbances does not have strong disturbances as would be expected so close to coalescence.

\subsection{SFR as a function of merger time}\label{sect:results:sfr}
\begin{figure}
	\resizebox{\hsize}{!}{\includegraphics{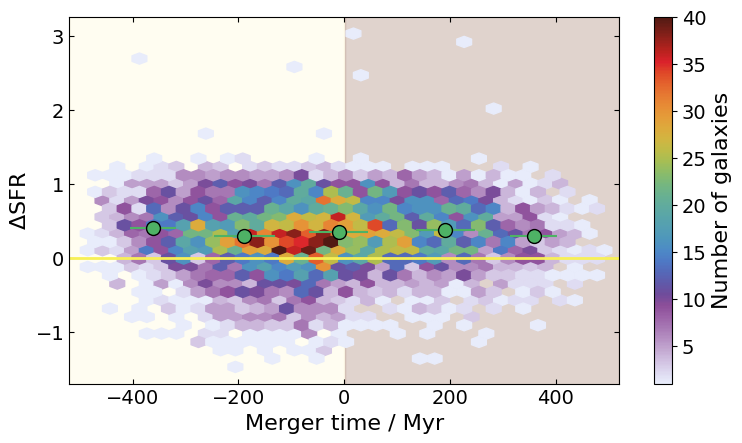}}
	\caption{$\Delta$SFR as a function of merger time. The green circles show the $\mu$ from the Gaussian fitting of the mergers. The errors on these points are the standard deviation of the merger times in the merger time bin (x axis) and the error in $\mu$ from the Gaussian fitting combined with bootstrapped $\Delta$SFR uncertainty (y axis). The colour corresponds to the number density from low (purple) to high (red). The light shaded region indicates where pre-mergers should lie and the dark shaded region indicates where post-mergers should lie. The yellow horizontal line indicates $\Delta\mathrm{SFR} = 0$.}
	\label{fig:results:SFRvsTime}
\end{figure}

We determined the change in the SFR of the mergers compared to the non-mergers by subtracting the median log(SFR/M$_\odot$yr$^{-1}$) of the ten non-merging control galaxies from the log(SFR/M$_\odot$yr$^{-1}$) of the merging system. That is to say,
\begin{equation}
	\Delta \mathrm{SFR} = \log\bigg(\frac{\mathrm{SFR}_{mgr}}{M_{\odot}yr^{-1}}\bigg) - median\bigg[ \log\bigg(\frac{\mathrm{SFR}_{nmg}}{M_{\odot}yr^{-1}}\bigg) \bigg],
\end{equation}
where $\Delta$SFR is SFR enhancement, SFR$_{mgr}$ is the SFR of the merging galaxy and SFR$_{nmg}$ is the SFR of a control galaxy. The mean (median) uncertainty on $\Delta$SFR, derived from the uncertainties in the SFR, is 0.60 (0.59).We can then plot the $\Delta$SFR as a function of merger time in Fig. \ref{fig:results:SFRvsTime}. As can be seen, the merging galaxies appear to primarily have an enhanced star formation rate ($\Delta$SFR $>$ 0).

\begin{figure}
	\resizebox{\hsize}{!}{\includegraphics{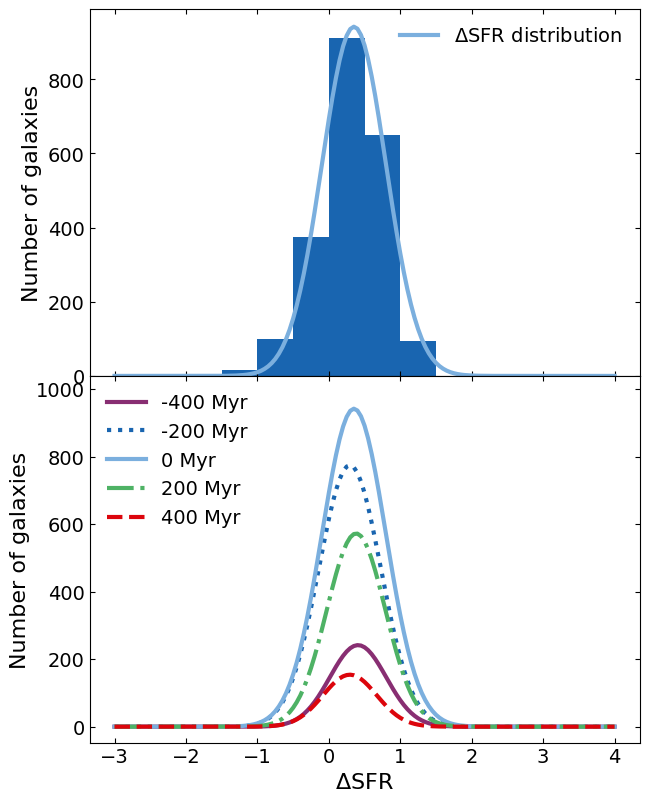}}
	\caption{Upper panel: Example of Gaussian fitting for the $-100$~Myr $\leq$ merger time $<$ 0~Myr bin. Lower panel: Gaussian fitting for all five merger time bins where the fit succeeded. The Gaussian distributions for the bins centred on $-400$~Myr (solid purple), $-200$~Myr (dotted dark blue), 0~Myr (solid light blue), 200~Myr (dash-dotted green), and 400~Myr (dashed red) are shown.}
	\label{fig:results:gaussian}
\end{figure}

To trace the change in $\Delta$SFR with merger time, we binned the galaxies by merger times with bins with a width of 200~Myr. For bins with more than ten galaxies, we attempted to fit a Gaussian distribution to their $\Delta$SFR with Scipy \texttt{curve\_fit} \citep{2020SciPy-NMeth}. An example of the Gaussian distribution can be found in Fig. \ref{fig:results:gaussian} (upper panel). We over-plot the mean of these Gaussian distributions on Fig. \ref{fig:results:SFRvsTime} as green circles.

The lower panel of Fig. \ref{fig:results:gaussian} shows the fitted Gaussian distributions for all merger time bins. As can be seen, the $\mu$ of the Gaussian distributions increases as the merger time increases from $-200$~Myr to $200$~Myr, before decreasing as it moves to $400$~Myr. The $\mu$ $\Delta$SFR take the values of 0.30$\pm$0.02, 0.35$\pm$0.02, 0.38$\pm$0.01, and 0.30$\pm$0.01, for the bins centred on $-200$~Myr, 0~Myr, $200$~Myr, and $400$~Myr, respectively. The $\mu$ $\Delta$SFR is highest in the bin centred on $-400$~Myr, at 0.41$\pm$0.01. Unsurprisingly, the amplitude of these distributions follows the distribution of merger times in Fig. \ref{fig:results:times}.

In short, from 300~Myr before a merger we see an increase in $\Delta$SFR as the merger approaches coalescence. This increase continues to increase into the post-merger regime before reducing again in late-stage mergers.

\subsection{Dependence on galaxy properties}
\subsubsection{Stellar mass}
\begin{figure}
	\resizebox{\hsize}{!}{\includegraphics{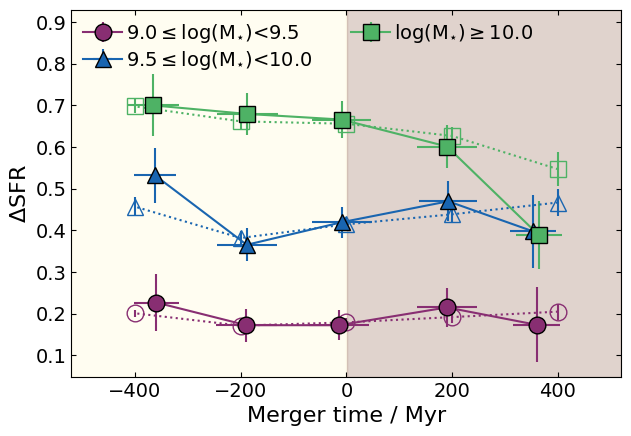}}
	\caption{$\Delta$SFR as a function of merger time for mergers binned by M$_{\star}$. The purple circles, blue triangles, and green squares show the $\mu$ from the Gaussian fitting of the binned merger times with masses of $9.0\leq \log(\mathrm{M}_{\star}/\mathrm{M}_{\odot})<9.5$, $9.5\leq \log(\mathrm{M}_{\star}/\mathrm{M}_{\odot})<10.0$, $\log(\mathrm{M}_{\star}/\mathrm{M}_{\odot}) \geq 10.0$, respectively. The errors on these points are the standard deviation of the merger times in the merger time bin (x axis) and the error in $\mu$ from the Gaussian fitting combined with bootstrapped $\Delta$SFR uncertainty (y axis). Empty markers show the results using re-sampled merger times (see text), and the y axis errors are the standard deviation of $\Delta$SFR from these resamplings. The light shaded region indicates where pre-mergers should lie and the dark shaded region indicates where post-mergers should lie.}
	\label{fig:results:mass}
\end{figure}

We binned the merging galaxies by M$_{\star}$ from log(M$_{\star}$/M$_{\odot}$) = 9.0 to log(M$_{\star}$/M$_{\odot}$) = 10.0 in steps of 0.5~log(M$_{\star}$/M$_{\odot}$) and one bin for $\log(\mathrm{M}_{\star}/\mathrm{M}_{\odot}) \geq 10.0$. For each mass bin, we fitted a Gaussian distribution, as is described in Sect. \ref{sect:results:sfr}, and show these results in Fig. \ref{fig:results:mass}.

Due to the large uncertainties of the merger times possibly impacting our results, we re-sampled the merger times of the KiDS galaxies. This was done by randomly selecting merger times from a Gaussian distribution centred on the predicted merger time from the CNN and with a standard deviation of 145~Myr (the median error) or the median merger time error in the merger time bin containing the galaxy (see Sect. \ref{sect:results:nn}), whichever is larger. Any galaxy whose new merger time is greater than 500~Myr or less than $-500$~Myr is then removed from further analysis. As with the CNN merger times, we fitted a Gaussian distribution, in the manner described in Sect. \ref{sect:results:sfr}. This resampling is continued until the average value of the means of the Gaussians for each merger time bin in every M$_{\star}$ bin does not change by more than 0.001. When this criteria is met, the $\Delta$SFR in each bin is assumed to have converged. These averages are presented in Fig. \ref{fig:results:mass} as empty markers.

There is a general trend of increasing $\Delta$SFR with increasing M$_{\star}$, with the exception of the high-mass mergers ($\log(\mathrm{M}_{\star}/\mathrm{M}_{\odot}) \geq 10.0$) at merger times of greater than $300$~Myr. The $\Delta$SFR of the high-mass mergers decreases as the merger time increases from $-500$~Myr to $500$~Myr. The decrease in $\Delta$SFR is gentler for mergers with merger times of less than $0$~Myr, with the changes between time bins within error, while there is a large drop between the bins centred on $200$~Myr and $400$~Myr. In this last merger time bin, centred on $400$~Myr, the $\Delta$SFR of the high-mass mergers is consistent with the $\Delta$SFR of the intermediate-mass mergers ($9.5 \leq \log(\mathrm{M}_{\star}/\mathrm{M}_{\odot}) < 10.0$). The intermediate-mass mergers' $\Delta$SFR also decreases from $-500$~Myr to $-200$~Myr but, unlike the high-mass mergers, the intermediate-mass ones show an increase in $\Delta$SFR as the merger time increases to $100$~Myr. Again, this mass range shows a decrease in $\Delta$SFR as the merger time continues to increase. The re-sampled merger times, on the other hand, continue to increase from $200$~Myr to the end of the merger time range. The low-mass mergers ($9.0 \leq \log(\mathrm{M}_{\star}/\mathrm{M}_{\odot}) < 9.5$) also see a decrease in $\Delta$SFR between $-500$~Myr and $-200$~Myr before remaining approximately constant to $0$~Myr. This is followed by an increase to $200$~Myr before falling again to the $400$~Myr merger time bin. The re-sampled low-mass mergers, however, have $\Delta$SFR that continues to rise from $200$~Myr to $500$~Myr.

\subsubsection{Local density}
\begin{figure}
	\resizebox{\hsize}{!}{\includegraphics{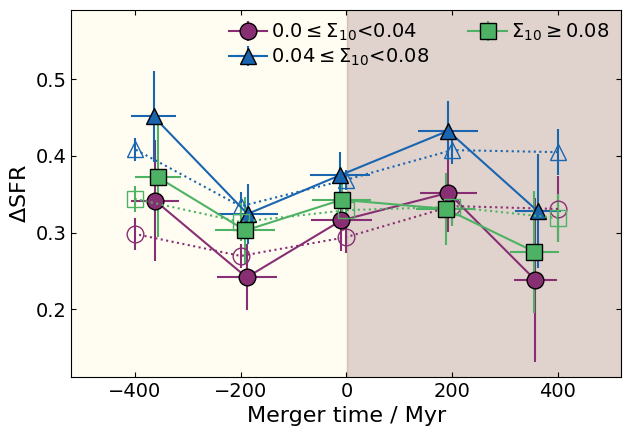}}
	\caption{$\Delta$SFR as a function of merger time for mergers binned by local density. The purple circles, blue triangles, and green squares show the $\mu$ from the Gaussian fitting of the binned merger times with $0.00\leq \Sigma_{10}<0.04$, $0.04\leq \Sigma_{10}<0.08$, and $\Sigma_{10}\geq 0.08$, respectively. The errors on these points are the standard deviation of the merger times in the merger time bin (x axis) and the error in $\mu$ from the Gaussian fitting combined with bootstrapped $\Delta$SFR uncertainty (y axis). Empty markers show the results using re-sampled merger times (see text), and the y axis errors are the standard deviation of $\Delta$SFR from these resamplings. The light shaded region indicates where pre-mergers should lie and the dark shaded region indicates where post-mergers should lie.}
	\label{fig:results:density}
\end{figure}

We binned the merging galaxies by $\Sigma_{10}$ into $0.00\leq \Sigma_{10}<0.04$, $0.04\leq \Sigma_{10}<0.08$, and $\Sigma_{10}\geq 0.08$. For each $\Sigma_{10}$ sample, we fitted a Gaussian distribution, as is described in Sect. \ref{sect:results:sfr}, and show these results in Fig. \ref{fig:results:density}. As with M$_{\star}$ above, we also resample the merger times until the average means of the Gaussians converges. These re-sampled averages are presented in Fig. \ref{fig:results:density} as empty markers.

The low-density ($0.00\leq \Sigma_{10}<0.04$) and intermediate-density ($0.04\leq \Sigma_{10}<0.08$) environments all show an increase in $\Delta$SFR as the galaxies move from merger times of $-200$~Myr to $200$~Myr, and then decrease again afterwards. The high-density ($\Sigma_{10}\geq 0.08$) environment has $\Delta$SFR increase between the $-200$~Myr and 0~Myr merger time bins followed by a decrease as the merger time becomes more positive, although these changes are within error. For all but the lowest densities, the $\Delta$SFR in the $-400$~Myr merger time bin is larger than at any other merger time, but again is consistent with later times within error. For the re-sampled merger times, this continues to hold. The intermediate-density environment typically has higher $\Delta$SFR than the other environments, being consistent with the high-density environment at merger times of $\approx-200$~Myr and $\approx400$~Myr.

\section{Discussion}\label{sect:discussion}
\subsection{Neural network}\label{sect:discussion:nn}
As expected, the CNN performs worse than the CNN in \citetalias{2024A&A...687A..45P}, as well as all the other architectures used in \citetalias{2024A&A...687A..45P} (Swin Transformer \citep{liu2021Swin}, and autoencoder Residual Network 50 \citep[ResNet50;][]{2015arXiv151203385H}). This is likely due to a combination of the noise added to the IllustrisTNG images as well as the convolution with the PSF. The convolution will reduce or erase the smaller scale structures that may be useful for determining the merger time. The inclusion of observational noise will hide fainter features that were found to be useful for the CNN in \citetalias{2024A&A...687A..45P}. Our results are better than a number of the autoencoder latent space explorations made in \citetalias{2024A&A...687A..45P}, specifically Isomap \citep{2000Sci...290.2319T}, neighbourhood components analysis \citep{NIPS2004_42fe8808}, sparse random projection \citep{10.1145/1150402.1150436}, and truncated singular value decomposition \citep{doi:10.1137/090771806}.

While our MSE in this work is similar to the ResNet50 used in \citetalias{2024A&A...687A..45P}, 0.067 with the validation set here compared to 0.065 at validation with ResNet50, these values are misleading. The results with ResNet50 in \citetalias{2024A&A...687A..45P} tended to a merger time of $-250$~Myr. As can be seen in Fig. \ref{fig:result:test}, we do not suffer the same issue. We do see that the mergers with true merger times greater than 0~Myr have predicted merger times that tend towards $240$~Myr, but mergers with true merger times of less than 0~Myr have predicted merger times that are reasonably well reproduced. We find a Pearson correlation coefficient of 0.57 with a p-value of $1.87\times10^{-59}$. Thus there is correlation between the predicted and true merger times but this correlation is weak.

In the test data, our network find 78$\pm$2\% of the galaxies to be correctly predicted as having merger times greater than, or of less than 0~Myr, lower than the 81$\pm$1\% found in \citetalias{2024A&A...687A..45P}. This is primarily driven by the galaxies with true merger times of less than 0~Myr, with 81$\pm$2\% of these galaxies predicted to have merger times of less than 0~Myr. The galaxies with true merger times greater than 0~Myr have a slightly worse reproduction with 77$\pm$2\% having predicted merger times greater than 0~Myr. This is qualitatively in line with \citetalias{2024A&A...687A..45P}.

\subsection{KiDS merger time}\label{sect:discussion:kids}
The merger times of the KiDS galaxies predominantly have merger times of less than 0~Myr. This is not overly surprising. The training data used to identify the KiDS merging galaxies in this work was from the GAMA-KiDS Galaxy Zoo citizen science project \citep{2008MNRAS.389.1179L, 2009A&G....50e..12D, 2019AJ....158..103H, 2019A&A...631A..51P, 2024PASA...41..115H}. Galaxies with negative merger times (pre-merger galaxies) can be easier to visually identify than galaxies with positive merger times (post-merger galaxies), as two close galaxies help in identification. Thus, if the training sample in \citet{2019A&A...631A..51P} was biased towards pre-mergers, we would expect the merger sample to be biased towards mergers with positive merger times. Indeed, early Galaxy Zoo merger samples were biased towards pre-merger galaxies \citep{2010MNRAS.401.1043D}.

In this work, we identify 2373 mergers with positive merger times and 3524 mergers with negative merger times. With our network assigning 23$\pm$1\% of galaxies with true merger times greater than 0~Myr negative merger times, we would expect approximately 709 of the galaxies with negative merger time to actually have positive merger times, assuming no true positive merger time galaxies were assigned negative merger times. This would agree with the idea that our sample of KiDS merging galaxies is biased towards pre-merger galaxies. On the other hand, if we incorrectly assign 19$\pm$2\% of galaxies with true negative merger times to have positive merger times, we would expect approximately 827 of the mergers with negative merger times to actually have positive merger times. This again assumes that no galaxies were assigned positive merger times that actually had negative merger times.

For all mergers, the mean and mode merger times of the IllustrisTNG test set are higher than those of the KiDS galaxies. The median of the test set is $-14\pm188$~Myr compared to $-48\pm157$~Myr for the KiDS sample (where the errors are the median absolute deviation), while the means are $18\pm9$~Myr and $-34\pm2$~Myr (where errors are the standard error on the mean) for the test and KiDS galaxies. This would suggest that the merger times for the KiDS galaxies are not just a reproduction of the distribution found for the test data. The Kolmogorov–Smirnov test also rejects the hypothesis that the test and KiDS merger times, for the pre-mergers, are drawn from the same parent sample at level 0.001. As a result, we are confident that the merger times are determined individually for each KiDS galaxy and are not drawn from an underlying, unphysical distribution that may have been learnt by the CNN.

\subsection{SFR as a function of merger time}\label{sect:discussion:sfr}
The merger galaxies, traced by green circles in Fig \ref{fig:results:SFRvsTime}, see an increasing enhancement for mergers with negative merger times from $-200$~Myr before coalescence to the merger event at a merger time of 0~Myr. This period is likely to contain close passages of the merging galaxies. Close passages are known, in simulations, to cause an increase in the SFR in the interacting galaxies \citep[e.g.][]{2019MNRAS.485.1320M, 2023MNRAS.518.3261P}. As merging galaxies will not have the close passages at exactly the same time, the $\Delta$SFR in individual galaxies due to a close pass will occur at slightly different times. Between passages, we would expect the SFR of the merging galaxies to decrease \citep[e.g.][]{2019MNRAS.485.1320M}. We do not see any decrease in $\Delta$SFR, or a constant $\Delta$SFR, as the merger time approached 0~Myr. This is possibly due to some merging systems' SFR reducing slightly after the passage while other galaxies experience another passage or the beginning of coalescence which stimulated further $\Delta$SFR. As a result, the average SFR increase becomes larger in this period with some galaxies experiencing a slight decline in SFR while others experience another burst. However, our merger time uncertainty is large. Thus, if galaxy mergers all had a close passage at approximately the same time, any decrease in $\Delta$SFR after a close passage can be lost due to the incorrect merger time. Similarly, the closer pre-merging galaxies are together, the higher the SFR of these galaxies \citep{2008AJ....135.1877E, 2008MNRAS.385.1903L, 2011A&A...535A..60H, 2013MNRAS.433L..59P, 2024OJAp....7E.121E, 2024ApJ...965...60F, 2025MNRAS.538L..31F}, which supports our observed increase in $\Delta$SFR as the merger time approached 0~Myr.

In short, the trends that we see in this work see an increase in SFR from 200~Myr before a merger until at least 200~Myr after coalescence. This is followed by a drop in SFR. This trend is broadly consistent with what is expected from simulations \citep[e.g.][]{2013MNRAS.430.1901H, 2019MNRAS.485.1320M, 2019MNRAS.490.2139R, 2025ApJ...979....7L}.

The higher $\Delta$SFR for merger times of less than $-300$~Myr was unexpected: we would expect this merger time bin to have one of the lowest $\Delta$SFR values \citep[e.g.][]{2019MNRAS.485.1320M, 2024OJAp....7E.121E}. This may be a period where a majority of mergers are experiencing a close passage, which are known to increase the SFR in merging galaxies. However, looking at the separation between merging galaxies in IllustrisTNG, we see a median or mean separation for mergers between $-300$~Myr and $-500$~Myr before the merger that is larger than the median or mean separation for mergers closer to coalescence. We also see very few merging galaxies with $\Delta$SFR of less than 0 with merger times between $-500$~Myr and $-300$~Myr compared to between $-300$~Myr and 0~Myr. As $\Delta$SFR $< 0$ suggests a reduction in SFR, it may be that mergers closer to coalescence can suppress star formation for a minority of merging galaxies.

The $\Delta$SFR continues to increase moving from 0~Myr to $200$~Myr. This may be a lingering effect of a SFR burst triggered by the coalescence or this burst takes up to 300~Myr to manifest. Also likely, the increase in $\Delta$SFR may be caused by high SFR mergers with a true merger time of $\approx$0~Myr, or with negative merger times close to 0~Myr, being scattered to more positive merger times as the CNN miss-assigns their merger time. We note that post-merger galaxies have previously been found to show evidence of increased SFR comparable to pre-mergers \citep{2022MNRAS.514.3294B, 2023MNRAS.518.3261P}. It is expected that the SFR decreases with time after the merger event \citep[e.g.][]{2019MNRAS.485.1320M, 2022MNRAS.514.3294B, 2022MNRAS.517L..92E, 2023MNRAS.518.3261P}, although the SFR enhancement can remain for up to 500~Myr after coalescence \citep{2025MNRAS.538L..31F}. Our results also show evidence of this, with the merger population showing a lower $\Delta$SFR for galaxies 300~Myr to 500~Myr after the merger event.

\begin{figure}
	\resizebox{\hsize}{!}{\includegraphics{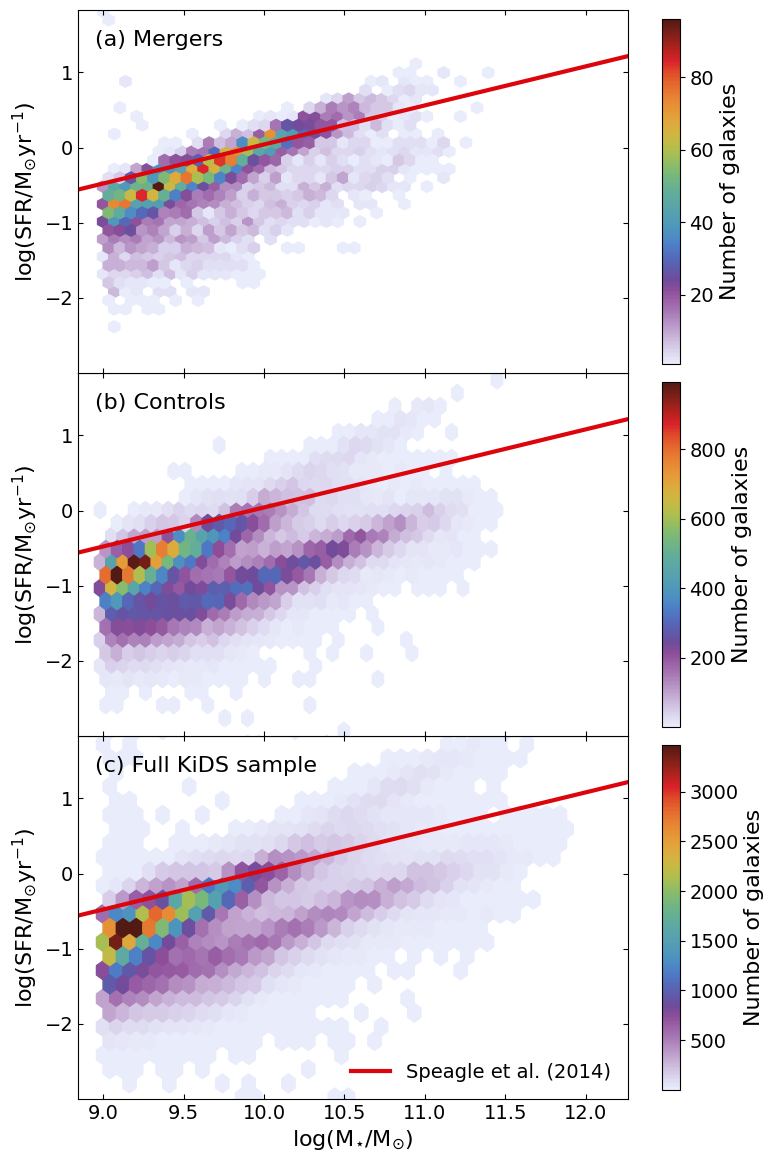}}
	\caption{SFR as a function of M$_{\star}$ for KiDS galaxies. Panel a shows the position of the merger sample, panel b shows the controls, and  panel c shows our entire sample of KiDS galaxies. The mergers lie on the MS, with a small number below.}
	\label{fig:discussion:MS}
\end{figure}

If the merging galaxies are plotted in the SFR-M$_{\star}$ plane, such as in Fig \ref{fig:discussion:MS}a, we find the main sequence of star-forming galaxies \citep[MS; e.g.][]{2007ApJ...660L..43N, 2014ApJS..214...15S, 2023MNRAS.519.1526P} and a small quiescent population. The control galaxies also lie across the entire SFR-M$_{\star}$ plane, as shown in Fig. \ref{fig:discussion:MS}b, as well as our entire KiDS galaxy sample, shown in Fig. \ref{fig:discussion:MS}c. We over-plot the \citet{2014ApJS..214...15S} compilation MS in Fig. \ref{fig:discussion:MS} to help identify the location of the MS. We note that the slope of the \citet{2014ApJS..214...15S} MS appears to be lower than the slope of the galaxies in this work. However, the slope of the MS is known to be sensitive to the star-forming galaxy selection method \citep{2023A&A...679A..35P}, and we make no selection for star-forming galaxies in Fig. \ref{fig:discussion:MS}. The low redshift ($z < 0.2$) slopes used in \citet{2014ApJS..214...15S} to derive the compilation MS are also highly scattered, ranging from 0.35 to 1. Thus a disagreement in the slope of the MS is not unexpected.

We find the enhancement of the merger population to be a factor of $1.97\pm0.04$ to $2.56\pm0.04$ larger than the control sample, depending on the merger time. This is consistent with or slightly higher than the enhancement of a factor of two found in \citet{2013MNRAS.433L..59P} and \citet{2025MNRAS.538L..31F}. However, \citet{2013MNRAS.433L..59P} and \citet{2025MNRAS.538L..31F} both select their merging galaxies and their control galaxies to be star-forming. We do not make such a selection and so we compare the SFR of our mergers to control samples that can contain star-forming and quiescent galaxies. As a result, our enhancement may be larger due to the median SFR of our matched control galaxies being lower than the average SFR of the matched control galaxies in \citet{2013MNRAS.433L..59P} and \citet{2025MNRAS.538L..31F}.

The SFRs used in this work estimate the instantaneous SFR, so should be sensitive to star formation over the last 1~Myr \citep{2019A&A...622A.103B}. However, spectral energy distribution (SED) derived SFRs are typically assumed to be sensitive to stars formed in the last 100~Myr \citep{2016ApJ...820L...1K}. Using this more conservative estimate, the SFR merger time sensitivity is smaller than our mean error at all merger times, and is smaller than the median error of true merger times greater than $-300$~Myr.

\subsection{Confirmation that mergers cause increase in $\Delta$SFR}
\begin{figure}
	\resizebox{\hsize}{!}{\includegraphics{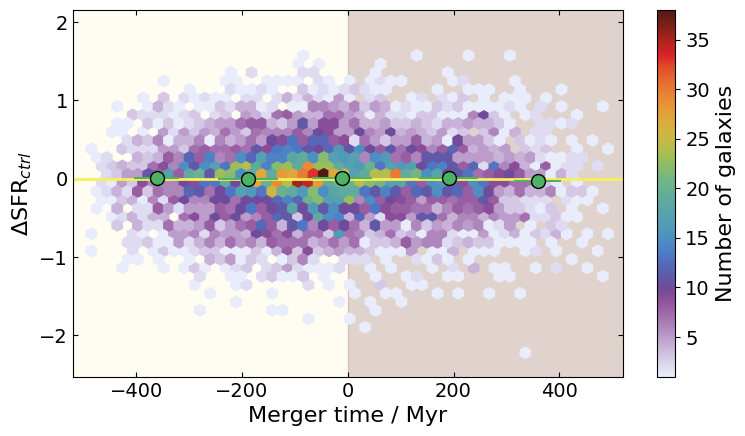}}
	\caption{SFR enhancement as a function of merger time for the control galaxies. The green circles show the $\mu$ from the Gaussian fitting of the mergers. The errors on these points are the standard deviation of the merger times in the merger time bin (x axis) and the error in $\mu$ from the Gaussian fitting combined with bootstrapped $\Delta$SFR uncertainty. The colour corresponds to the number density from low (purple) to high (red). The light shaded region indicates where pre-mergers should lie and the dark shaded region indicates where post-mergers should lie. The yellow horizontal line indicates $\Delta\mathrm{SFR}_{ctrl} = 0$}
	\label{fig:discussion:control}
\end{figure}

To ensure that the enhancement in the SFR seen is caused by mergers, we repeated the calculation of SFR enhancement using only our non-merging control galaxies. For each sample of ten non-merging control galaxies, we randomly selected one and compared its SFR to the other nine control galaxies. This was done in the same way as we did for the merging galaxies: the log(SFR/M$_{\odot}$yr$^{-1}$) of one control galaxy was subtracted from the median log(SFR/M$_{\odot}$yr$^{-1}$) of the other nine control galaxies. That is to say,
\begin{equation}
	\Delta \mathrm{SFR}_{ctrl} = \log\bigg(\frac{\mathrm{SFR}_{r}}{M_{\odot}yr^{-1}}\bigg) - median \bigg[ \log\bigg(\frac{\mathrm{SFR}_{c}}{M_{\odot}yr^{-1}}\bigg) \bigg],
\end{equation}
where $\Delta$SFR$_{ctrl}$ is the SFR enhancement of the randomly selected control galaxy,  SFR$_{r}$ is the SFR of the randomly selected control galaxy and SFR$_{c}$ is the SFR of a the other nine control galaxies. The mean (median) uncertainty on $\Delta$SFR$_{ctrl}$, derived from the uncertainties in SFR, is 0.65 (0.66) We then use the merger time of the merging galaxy associated with each control sample to create Fig. \ref{fig:discussion:control}, over-plotting the $\mu$ of fitting a Gaussian distribution to the $\Delta$SFR in merger time bins as in Sect. \ref{sect:results:sfr} and in Fig. \ref{fig:results:SFRvsTime}. As can be seen, there is no indication of $\Delta$SFR being larger than 0. Indeed, we find that the mean $\Delta$SFR$_{ctrl}$ in Fig. \ref{fig:discussion:control} is $-0.02\pm0.47$, the median is $0.00\pm0.36$, and the Gaussian $\mu$ (equivalent to the mode) is $0.00\pm0.02$. All these values are consistent with zero and so the enhancement seen in Fig. \ref{fig:results:SFRvsTime} for the merging galaxies is a result of the mergers causing changes in the SFR.

\subsection{Merger time uncertainty}\label{sect:discussion:sigmat}

The merger times found in this work are highly uncertain. To determine how this uncertainty impacts our results, we resample the merger times of the KiDS galaxies from a Gaussian distribution centred on the merger time from the CNN and with a standard deviation of 145~Myr (the median error) or the median merger time error in the merger time bin containing the galaxy (see Sect. \ref{sect:results:nn}), whichever is larger. Any galaxy whose new merger time is greater than 500~Myr or less than $-500$~Myr is then removed from further analysis as our network is not trained to produce such merger times. We also do not recover the merging galaxies who would have merger time estimates of less than $-500$~Myr or greater than 500~Myr but whose resampling would scatter them back into the $-500$~Myr to 500~Myr range. We binned the merging galaxies into bins of width 200~Myr and fitted a Gaussian distribution, as is described in Sect. \ref{sect:results:sfr}. This allowed us to trace the merger population with re-sampled merger times.

\begin{figure}
	\resizebox{\hsize}{!}{\includegraphics{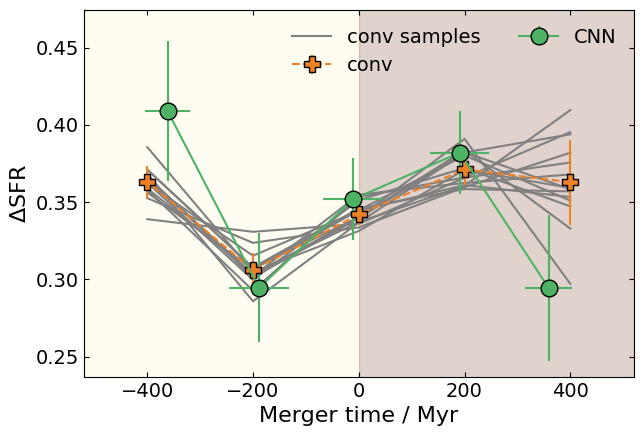}}
	\caption{$\Delta$SFR as a function of merger time. The green circles and orange pluses show the $\mu$ from the Gaussian fitting of the mergers using the CNN merger times and conv merger times, respectively. The errors on these points are the standard deviation of the merger times in the merger time bin (x axis) and the error in $\mu$ from the Gaussian fitting combined with bootstrapped $\Delta$SFR uncertainty, for CNN merger time, or the standard deviation of $\Delta$SFR, for conv (y axis). The grey lines show $\Delta$SFR from the merger time re-samplings used to determine conv. Errors for these lines are omitted for clarity. The light shaded region indicates where pre-mergers should lie and the dark shaded region indicates where post-mergers should lie.}
	\label{fig:discussion:resample}
\end{figure}

We continued to resample the merger times until the mean $\Delta$SFR converged. This was done by continuing to resample the merger time for each merger galaxy from a Gaussian distribution centred on the merger time from the CNN and with a standard deviation of 145~Myr (the median error) or the median merger time error in the merger time bin containing the galaxy (see Sect. \ref{sect:results:nn} and Table \ref{table:results:errors}), whichever is larger. Again we removed any galaxy whose re-sampled merger time is less than $-500$~Myr or greater than 500~Myr. After each resampling, we again binned the merging galaxies into bins of width 200~Myr and fitted a Gaussian distribution. The five merger time bins provide five averages. If each of these five average values have not changed by more than 0.001 and the sum of these changes is not more than 0.005, the average is assumed to have converged. This was achieved after 13 re-samplings. These converged average $\Delta$SFR are hereafter referred to as conv and are presented in Fig. \ref{fig:discussion:resample} as orange pluses. The re-samplings used to derive conv are shown as grey lines in Fig. \ref{fig:discussion:resample}.

For the conv resampling, $\Delta$SFR increases as merger times move from $-200$~Myr to $200$~Myr, meaning that mergers that should have recently completed a merger have higher $\Delta$SFR than mergers that are coalescing or approaching coalescence. Galaxies with merger times greater than $300$~Myr appear to have lower $\Delta$SFR than mergers with smaller merger times, but this drop is within uncertainty of the bin centred on $400$~Myr so is not robust. At the other end of the time range, we again see the conv resampling has higher $\Delta$SFR for merger times of less than $-300$~Myr when compared to galaxies with merger times between $-300$~Myr and $0$~Myr.

With the increase in $\Delta$SFR for the conv and CNN merger times, we can be cautiously optimistic that we are seeing in enhancement in the SFR as the mergers approach coalescence. As all the re-samplings and the CNN see a continued $\Delta$SFR enhancement into the early post-merger phase, it is likely that early post-mergers continue to have increasing SFR compared to mergers with merger times of less than 0~Myr.

The large decrease in $\Delta$SFR moving from the $200$~Myr to the $400$~Myr bins with the CNN merger times is not strongly seen in the conv data and only seen in one of the re-samplings. Therefore, we believe we are only seeing weak evidence of this decrease in $\Delta$SFR. This weak evidence of a decreasing $\Delta$SFR is likely due to the poor merger time accuracy for galaxies with merger times greater than 0~Myr, with galaxies in the 200~Myr and 400~Myr bins being easily confused by the neural network (Fig. \ref{fig:result:test}).

One of the re-samplings in Fig. \ref{fig:discussion:resample} only shows a very slight drop in $\Delta$SFR between the bins centred on $-400$~Myr and $-200$~Myr. This drop is within the errors of these points for this resampling (not shown in Fig. \ref{fig:discussion:resample}). This hints that the unexpectedly large $\Delta$SFR for merger times of less than $-300$~Myr may also be a result of our merger time uncertainty. However, as this only appears in a single resampling, this is not a very strong argument for why we see this increase with the CNN merger times.

\subsection{Dependence on galaxy properties}\label{sect:discussion:properties}

\subsubsection{Stellar mass}\label{sect:discussion:properties:mass}

High-mass galaxies are more likely to be gas-poor, and thus high mass major mergers are more likely to have less gas available to form new stars \citep{2014MNRAS.444.3986R}. This may explain why the $\Delta$SFR of our high-mass galaxies ($\log(\mathrm{M}_{\star}/\mathrm{M}_{\odot}) \geq 10.0$) drops to become consistent with the intermediate-mass mergers ($9.5 \leq \log(\mathrm{M}_{\star}/\mathrm{M}_{\odot}) < 10.0$) for the most positive merger times in Fig. \ref{fig:results:mass}. However, this does not explain why the $\Delta$SFR for the high-mass galaxies is otherwise higher than the other mass bins. It is possible, instead, that the gas is depleted faster in high-mass galaxy mergers compared to the other mass bins: higher $\Delta$SFR implies a higher specific SFR and hence faster gas depletion. If the SFR is triggered sooner in a high-mass merger, this would explain why $\Delta$SFR decreases slowly for mergers with merger times of less than 0~Myr. This rapid depletion of gas would then result in a quicker decrease in $\Delta$SFR for high-mass galaxies than intermediate- or low-mass galaxies. If this is what is happening, high-mass mergers are not initially gas-poor (i.e. the high-mass mergers were star-forming before they merged) but the gas is depleted quickly compared to lower mass mergers. $\Delta$SFR for the intermediate-mass galaxies also reduces faster as the merger time increases from 200~Myr to 500~Myr compared to the low-mass galaxies, supporting the idea that the higher the mass of a merger, the faster it depletes the gas available for star formation.

$\Delta$SFR increases with M$_{\star}$, as seen in Fig. \ref{fig:results:mass}, at least for the population with merger times of less than $300$~Myr. We note that the $\Delta$SFR compares galaxies of similar M$_{\star}$, so this increase with stellar mass is not caused by the correlation between M$_{\star}$ and SFR. This would suggest that there is more gas available to form stars as the galaxies undergoing a merger become larger.

However, the control sample for high-mass mergers mainly consists of quiescent galaxies (Fig. \ref{fig:discussion:MS}). Here, we define a star-forming galaxy to have a specific SFR (SFR/M$_{\star}$) greater than $-10.5$~yr$^{-1}$, determined through visual inspection of Fig \ref{fig:discussion:MS}. With this definition, only 23$\pm$1\% of the high-mass control galaxies are star-forming while 72$\pm$1\% of the high-mass mergers are star-forming. This compares to the low-mass sample having 78$\pm$1\% star-forming controls and 85$\pm$1\% star-forming mergers while the intermediate-mass sample has 52$\pm$1\% of the controls being star-forming and 85$\pm$1\% of the mergers being star-forming. As a result, the $\Delta$SFR increase with M$_{\star}$ may predominantly be a result of the higher fraction of quiescent control galaxies as the merging galaxies' M$_{\star}$ increases. If we compare this to the entire sample of KiDS galaxies, 69$\pm$1\% of low-mass galaxies, 51$\pm$1\% of intermediate-mass galaxies, and 24$\pm$1\% of high-mass galaxies are star-forming. This increasing fraction of quiescent galaxies with M$_{\star}$ in the control sample is therefore a result of the increasing fraction of quiescent galaxies with M$_{\star}$ in the entire sample of KiDS galaxies. This increase in the quiescent fraction with M$_{\star}$ is often seen in the literature \citep[e.g.][]{2010ApJ...721..193P, 2023A&A...677A.184W}.

\subsubsection{Local density}
High-density environments contain more gas-poor galaxies than lower-density environments. Thus, we expect more dry mergers, and hence lower $\Delta$SFR mergers \citep{2010ApJ...718.1158L, 2024RAA....24e5005H}. In Fig. \ref{fig:results:density} we typically see the high-density environments ($\Sigma_{10}\geq 0.08$) have lower $\Delta$SFR than the intermediate-density environments ($0.04\leq \Sigma_{10}<0.08$), which supports this notion. However, the low-density environments ($0.00\leq \Sigma_{10}<0.04$) are typically found to have lower $\Delta$SFR than intermediate- and high-density environments. This is inconsistent, or at best marginally consistent, with studies finding higher-density environments containing more gas-poor galaxies, or in general contain less gas \citep{2010ApJ...718.1158L}. It is not clear why this should be the case as we would expect the lower-density environments to have more gas available to form stars.

The general trend for low- and intermediate-density environments are the same as the entire population: $\Delta$SFR drops from the merger time bin centred on $-400$~Myr to the $-200$~Myr before increasing to $200$~Myr and then falling into the final merger time bin. The high-density environments differ from this trend after 0~Myr, as this environment only sees a decrease in $\Delta$SFR from this merger time. As is discussed in Sect. \ref{sect:discussion:sigmat}, this difference may be due to merger time uncertainty.

However, the disagreement between environments is different in the converged, re-sampled merger times in Fig. \ref{fig:results:density} (empty markers). All environments remain approximately constant in $\Delta$SFR between the 200~Myr and 400~Myr merger time bins. These differences suggest that the strength of the trends seen with the CNN merger times are at least partially a result of the merger time uncertainty.

\section{Conclusions}\label{sect:conclusion}
In this work, we trained a CNN on images of merging galaxies from IllustrisTNG with known merger times with the intent to apply this CNN to observed merging galaxies from KiDS. The images of the IllustrisTNG galaxies were edited to appear like KiDS observations. The merging galaxies were selected from TNG100-1 to have undergone a merger in the last 500~Myr or will undergo a merger in the next 500~Myr. The trained CNN has a mean error of 182~Myr and a median error of 145~Myr.

The CNN was applied to 5897 merging galaxies from KiDS, providing merger times for a large sample of observed galaxies for the first time. The majority of the KiDS mergers (3524) were found to have merger time consistent with pre-mergers. This was consistent with our expectation that the KiDS sample is biased towards pre-mergers, as the KiDS mergers were selected with a CNN from \citet{2019A&A...631A..51P} that was trained on a merger sample that was likely biased towards pre-mergers. The merging galaxies were matched to non-merging counterparts with similar redshift, M$_{\star}$, and local density ($\Sigma_{10}$). The SFR of the merging systems were compared to the average SFR of their matched counterparts.

Our merging galaxies have increasing SFR activity between 300~Myr before the merger event and coalescence (merger times of 0~Myr). We suggest that this increase is due to closer proximity of the merging galaxies as well as close passages as the merger time approached 0~Myr. This is followed by a further increase in the SFR remaining until at least 200~Myr after the merger event. We also see an indication that this is followed by a decrease in the SFR activity as the time since a merger increases further. This decrease in $\Delta$SFR for late-stage mergers is difficult to be confident in due to the larger merger time uncertainties for late-stage mergers. There is also a decrease in $\Delta$SFR from $\approx$400~Myr before the merger event and $\approx$200~Myr before. The reason for this is unclear but there is evidence that it may be a result of the low accuracy in merger time estimation.

When binning by M$_{\star}$, we see that the larger the merging galaxy's M$_{\star}$, the greater the SFR enhancement. We generally find that pre-mergers with higher M$_{\star}$ have higher $\Delta$SFR. However, the most massive enhanced merging galaxies have $\Delta$SFR that decreases faster after the merger is complete compared to the lower mass galaxies. When binned by $\Sigma_{10}$, we find that the lowest density environments have lower $\Delta$SFR mergers than higher-density environments and intermediate density environments have higher $\Delta$SFR mergers than other environments. However, this trend is inconsistent, suggesting that density does not play a major roll in the change in SFR in galaxies due to mergers.

The merger times produced in this work are not highly accurate. However, they show that it is possible to determine the time before or after a merger event for real galaxies, something not previously managed on such a large scale. With these times, we can then study how galaxies change during the merger sequence from a statistical standpoint. Future work will use a target observed data set with deeper observations, to better show faint structures, with higher spatial resolution, to better show fine structures. For example, targeting Hyper-Suprime Cam - Subaru Strategic Program \citep{2018PASJ...70S...4A} data will provide deeper imaging than KiDS at a similar spatial resolution, Vera C. Rubin \citep{2019ApJ...873..111I} data will provide deeper imaging at a similar spatial resolution to KiDS but cover most of the Southern sky, while targeting James Web Space Telescope \citep{2006SSRv..123..485G} data will provide both deeper images and greater resolution at the expense of a smaller area of sky coverage.

\section{Data availability}
Table \ref{table:results:times} is only available in electronic form at the CDS via anonymous ftp to cdsarc.u-strasbg.fr (130.79.128.5) or via http://cdsweb.u-strasbg.fr/cgi-bin/qcat?J/A+A/

\begin{acknowledgements}
	We would like to thank the referee for their thorough and thoughtful comments that helped improve the quality and clarity of this paper.
	
	W.J.P. has been supported by the Polish National Science Center projects UMO-2020/37/B/ST9/00466 and UMO-2023/51/D/ST9/00147.
	
	L.W. and B.M-B acknowledge funding from the project `Clash of the Titans: de-ciphering the enigmatic role of cosmic collisions' (with project number VI.Vidi.193.113 of the research programme Vidi which is (partly) financed by the Dutch Research Council (NWO).
	
	L.E.S. was supported by the Estonian Ministry of Education and Research (grant TK202), Estonian Research Council grant (PRG1006), and the European Union's Horizon Europe research and innovation programme (EXCOSM, grant No. 101159513).
	
	The IllustrisTNG simulations were undertaken with compute time awarded by the Gauss Centre for Supercomputing (GCS) under GCS Large-Scale Projects GCS-ILLU and GCS-DWAR on the GCS share of the supercomputer Hazel Hen at the High Performance Computing Center Stuttgart (HLRS), as well as on the machines of the Max Planck Computing and Data Facility (MPCDF) in Garching, Germany.
	
	Based on observations made with ESO Telescopes at the La Silla Paranal Observatory under programme IDs 177.A-3016, 177.A-3017, 177.A-3018 and 179.A-2004, and on data products produced by the KiDS consortium. The KiDS production team acknowledges support from: Deutsche Forschungsgemeinschaft, ERC, NOVA and NWO-M grants; Target; the University of Padova, and the University Federico II (Naples).
	
	This work made use of Astropy:\footnote{\url{http://www.astropy.org}} a community-developed core Python package and an ecosystem of tools and resources for astronomy \citep{2013A&A...558A..33A, 2018AJ....156..123A, 2022ApJ...935..167A}.
	
	This research made use of Photutils, an Astropy package for detection and photometry of astronomical sources \citep{larry_bradley_2024_10671725}.
\end{acknowledgements}

\bibliographystyle{aa}
\bibliography{aa53228-24}

\begin{appendix}

\onecolumn
\section{CIGALE setup}\label{app:cigale}
Here we present our CIGALE setup used in this work in Table \ref{tab:cigaletable}.

\begin{table*}[ht!]\label{tab:cigaletable}
	\caption{Parameters used for the CIGALE model SEDs used to estimate M$_{\star}$ and SFR. All ages and times are in Myr.}
	\centering
	\begin{tabular}{l c c}
		\hline
		\hline
		Parameter & Value & Description\\
		\hline
		\multicolumn{3}{c}{Star Formation History} \\
		\hline
		$\tau_{\mathrm{main}}$ & 1000, 1800, 3000, 5000, 7000 & e-folding time (main)\\
		$\tau_{\mathrm{burst}}$ & 9, 13 & e-folding time (burst) \\
		$f_{\mathrm{burst}}$ & 0.00, 0.10, 0.20, 0.30 & Burst mass fraction\\
		Age & 1000, 1500, 2000, 2500, 3000, 3500, & Population age (main)\\
		& 4000, 4500, 5000, 5500, 6000, 6500, & \\
		& 7000, 7500, 8000, 8500, 9000, 9500, & \\
		& 10000, 10500, 11000, 11500, 12000, & \\
		& 13000 & \\
		Burst Age & 1, 10, 30, 100, 300 & Population age (burst)\\
		& & \\
		\hline
		\multicolumn{3}{c}{Stellar Emission \citep{2003MNRAS.344.1000B}}\\
		\hline
		IMF & \citet{2003PASP..115..763C} & Initial Mass Function\\
		$Z$ & 0.02 & Metallicity (0.02 is Solar)\\
		Separation Age & 10 & Separation between young and old stellar populations\\
		& & \\
		\hline
		\multicolumn{3}{c}{Dust Attenuation \citep[Modified ][]{2000ApJ...539..718C}}\\
		\hline
		A$_\mathrm{V}^{\mathrm{BC}}$ & 0.3, 1.2, 2.3, 3.3, 3.8 & V-band attenuation of the birth clouds\\
		$\mu$ & 0.23, 0.33, 0.44, 0.5 & Av$_{ISM}$ / (Av$_{BC}$+Av$_{ISM}$)\\
		Slope$_{\mathrm{BC}}$ & -0.7 & Birth cloud attenuation power law slope\\
		Slope$_{\mathrm{ISM}}$ & -0.7 & ISM attenuation power law slope\\
		& & \\
		\hline
		\multicolumn{3}{c}{Dust Emission \citep{2014ApJ...780..172D}}\\
		\hline
		$q_{\mathrm{PAH}}$ & 0.47, 1.12, 2.50, 3.9 & Mass fraction of PAH \\
		$U_{\mathrm{min}}$ & 5.0, 10.0, 25.0 & Minimum scaling factor of the radiation field intensity\\
		$\alpha$ & 2.0 & Dust power law slope\\
		$\gamma$ & 0.02 & Illuminated fraction\\
		& & \\
		\hline
		\multicolumn{3}{c}{AGN \citep{2006MNRAS.366..767F}}\\
		\hline
		$r_{ratio}$ & 60 & Ratio of the maximum to minimum radii of the dust torus\\
		$\tau$ & 1.0, 6.0 & Optical depth at 9.7 microns\\
		$\beta$ & -0.5 & Beta\\
		$\gamma$ & 0.0 & Gamma\\
		opening$_{\mathrm{angle}}$ & 100 & Opening angle of the dust torus\\
		psy & 0.001, 89.990 & Angle between equatorial axis and line of sight\\
		disk$_{\mathrm{type}}$ & 1 & Disk spectrum \citep[1 is ][]{2005AA...437..861S} \\
		$\delta$ & -0.36 & Power-law of index $\delta$ modifying the optical slop of the disk.\\
		fracAGN & 0.0, 0.1, 0.3, 0.5, 0.7 & AGN fraction\\
		lambda$_{\mathrm{fracAGN}}$ & 0/0 & Wavelength range in microns where to compute the AGN fraction\\
		law & 0 & Extinction law of the polar dust (0 = Small Magellanic Cloud)\\
		EBV & 0.03 & E(B-V) for the extinction in the polar direction in magnitudes\\
		temperature & 100 & Temperature of the polar dust in K\\
		emissivity & 1.6 & Emissivity index of the polar dust\\
		& & \\
		\hline
		
	\end{tabular}
\end{table*}

\twocolumn
\FloatBarrier
\section{Comparison of SFR and M$_{\star}$ to GAMA}\label{app:sfr}
Here we compare our SFR and M$_{\star}$ values derived through SED fitting described in Sect. \ref{sect:data:sfr}. We match KiDS galaxies to galaxies found in the fourth data release of the Galaxy and Mass Assembly survey \citep[GAMA;][]{2009A&G....50e..12D, 2022MNRAS.513..439D}. We match all KiDS galaxies with $z < 0.15$ to GAMA galaxies that are within 2~arcsec and have redshifts within 0.01. In total, we match to 5212 GAMA galaxies.

\begin{figure}
	\resizebox{\hsize}{!}{\includegraphics{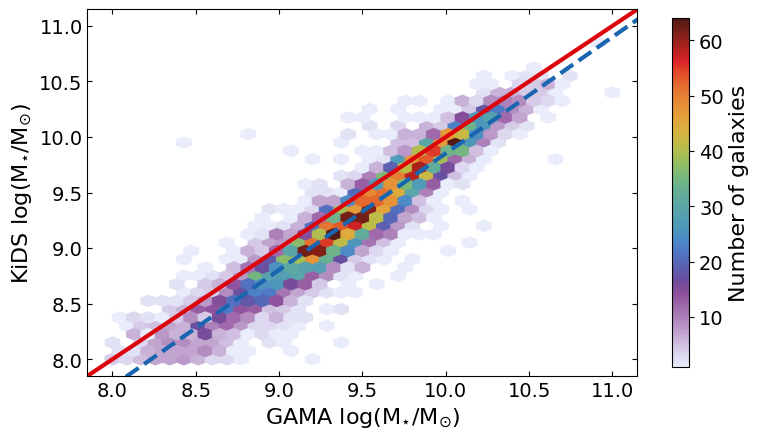}}
	\caption{Our CIGALE derived M$_{\star}$ for KiDS galaxies as a function of M$_{\star}$ for matched GAMA galaxies. The colour corresponds to the number density from low (purple) to high (red). The solid red line indicates the one-to-one relation, while the dashed blue line indicates our fitted relation.}
	\label{fig:app:mass}
\end{figure}

For M$_{\star}$, we compared our values to the values derived by GAMA using SED fitting to the Lambdar photometry \citep{2011MNRAS.418.1587T, 2016MNRAS.460..765W}, presenting the distributions in Fig. \ref{fig:app:mass}. We fitted a straight line to the relation between the GAMA log(M$_{\star}$/M$_{\odot}$) and our KiDS log(M$_{\star}$/M$_{\odot}$). We found a slope of $1.05\pm0.01$ and an offset of $-0.20\pm0.01$. This suggests that our M$_{\star}$ are systematically lower than the GAMA M$_{\star}$ by 0.2~dex. Using our CIGALE setup with the Lambdar data from GAMA, we find a very small offset. Again fitting a straight line to the relation between the GAMA log(M$_{\star}$/M$_{\odot}$) and the log(M$_{\star}$/M$_{\odot}$) found using our CIGALE setup and the GAMA Lambdar photometry, we find a slope of 0.99$\pm$0.01 and an offset of $-0.03\pm0.01$. This suggests that the offset we observe is a result of the limited data used for fitting: here the 9 KiDS+VIKINGS optical and NIR bands compared to the 21 ultraviolet (UV) to far-infrared (FIR) bands used in GAMA.

\begin{figure}
	\resizebox{\hsize}{!}{\includegraphics{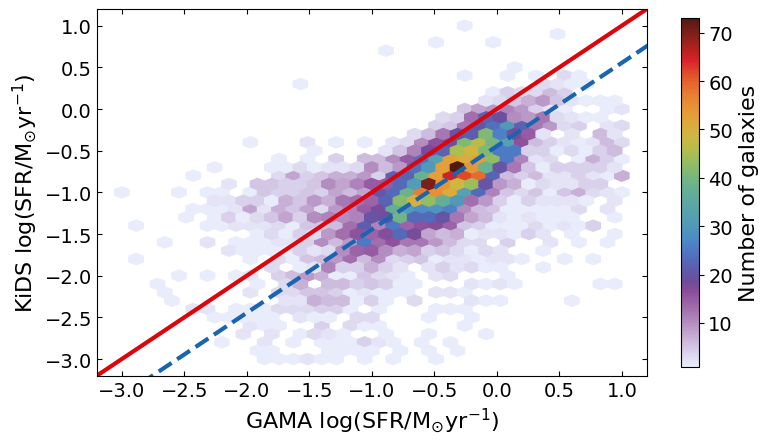}}
	\caption{Our CIGALE derived SFR for KiDS galaxies as a function of SFR for matched GAMA galaxies derived from H$\alpha$ observations. The colour corresponds to the number density from low (purple) to high (red). The solid red line indicates the one-to-one relation, while the dashed blue line indicates our fitted relation with unity slope.}
	\label{fig:app:sfr}
\end{figure}

For SFR, we compared our values to values derived from GAMA H$\alpha$ observations using the Kennicutt relation \citep{1983ApJ...272...54K}. We present this comparison in Fig. \ref{fig:app:sfr}. Here we fit a straight line with a slope of unity to the galaxies where both log(SFR/M$_{\odot}$yr$^{-1}$) are between $-3$ and 1. We find a systematic offset with our CIGALE derived SFRs being 0.45$\pm$0.01~dex below the H${\alpha}$ SFRs. A disagreement was not unexpected as we do not have any UV or FIR data in our SED fitting, both of which improve SFR estimation. Again using our CIGALE setup to estimate the SFR with the GAMA Lambdar photometry, we find a smaller underestimation of $-0.13\pm0.01$~dex. This smaller offset again suggests that our underestimation is primarily due to the limited photometry available across the entire KiDS fields.

We also compared our KiDS results and the GAMA results for only the merging galaxies and only the non-merging galaxies. Fitting the log(SFR) of the 308 matched merging galaxies with a straight line with unity slope, we find an offset of $-0.64\pm$0.04. The 2787 matched non-mergers have an offset of $-0.35\pm$0.01. Thus, the mergers have a larger underestimation compared to the non-mergers, assuming the GAMA SFRs are correct. We note that the missing 2117 galaxies are not classified as mergers or non-mergers using our criteria in Sect. \ref{sect:data:mergers}.

Both M$_{\star}$ and SFR used in this work have a systematic underestimation compared to the results from GAMA. However, M$_{\star}$ is only used to select control galaxies while the SFRs are used in comparison to other SFRs which should alleviate the impact of constant systematic offsets seen here. We see larger underestimation for the merging galaxies compared to the non-merger galaxies, meaning $\Delta$SFR may be underestimated. As a result, we elect to not correct for these offsets in this work.

\end{appendix}

\end{document}